\documentclass[aps,prl,reprint,groupedaddress,longbibliography,nobalancelastpage]{revtex4-2}
\usepackage{amsmath,amssymb,amsthm,dcolumn}
\usepackage{graphicx}
\usepackage{xcolor}
\usepackage{lmodern}

\usepackage{hyperref}
\hypersetup{colorlinks,linkcolor={blue!90!black},citecolor={blue!90!black},urlcolor={blue!90!black}}
\usepackage{stmaryrd}
\usepackage{enumitem}
\usepackage{accents}
\usepackage{CJK}
\renewcommand{\vec}[1]{\boldsymbol{#1}}

\DeclareMathAlphabet{\mathbfsfit}{\encodingdefault}{\sfdefault}{bx}{sl}
\newcommand{\tens}[1]{\mathbfsfit{#1}}

\renewcommand{\eqref}[1]{\hyperref[#1]{(\ref*{#1})}}
\newcommand{\figref}[2]{[Fig.~\hyperref[#1]{\ref*{#1}(#2)}]}
\newcommand{\bfigref}[3]{[Fig.~\hyperref[#1]{\ref*{#1}(#2\textsubscript{#3})}]}
\newcommand{\figrefi}[2]{[Fig.~\hyperref[#1]{\ref*{#1}(#2)}, inset]}
\newcommand{\textfigref}[2]{Fig.~\hyperref[#1]{\ref*{#1}(#2)}}
\newcommand{\textfigureref}[2]{Figure~\hyperref[#1]{\ref*{#1}(#2)}}

\newcommand{\textwholefigref}[1]{Fig.~\ref{#1}}

\newcommand{\figrefp}[2]{\hyperref[#1]{\ref*{#1}(#2)}}

\DeclareMathAlphabet{\mathcalbf}{OMS}{cmsy}{b}{n}
\DeclareMathAlphabet{\mathbfsf}{\encodingdefault}{\sfdefault}{b}{n}
\DeclareMathOperator{\tr}{tr}
\DeclareMathOperator{\dev}{dev}

\DeclareFontEncoding{LGR}{}{}
\DeclareSymbolFont{sfgreek2}{LGR}{cmss}{bx}{it}
\DeclareMathSymbol{\teps}{\mathord}{sfgreek2}{`e}

\definecolor{linkcolor}{HTML}{223096}
\hypersetup{colorlinks,allcolors=linkcolor}

\bibpunct{\textcolor{linkcolor}{[}}{\textcolor{linkcolor}{]}}{\textcolor{linkcolor}{,}}{n}{}{;}
\renewcommand{\i}{\mathrm{i}}
\newcommand{\e}{\mathrm{e}}

\begin{document}

\begin{CJK*}{UTF8}{gbsn}

\title{Nonlinear isotropic odd elasticity}

\author{Shiheng Zhao (赵世恒)}
\affiliation{Max Planck Institute for the Physics of Complex Systems, N\"othnitzer Stra\ss e 38, 01187 Dresden, Germany}
\affiliation{\mbox{Max Planck Institute of Molecular Cell Biology and Genetics, Pfotenhauerstra\ss e 108, 01307 Dresden, Germany}}
\affiliation{Center for Systems Biology Dresden, Pfotenhauerstra\ss e 108, 01307 Dresden, Germany}
\author{Pierre A. Haas}
\email[Contact author: ]{haas@pks.mpg.de}
\affiliation{Max Planck Institute for the Physics of Complex Systems, N\"othnitzer Stra\ss e 38, 01187 Dresden, Germany}
\affiliation{\mbox{Max Planck Institute of Molecular Cell Biology and Genetics, Pfotenhauerstra\ss e 108, 01307 Dresden, Germany}}
\affiliation{Center for Systems Biology Dresden, Pfotenhauerstra\ss e 108, 01307 Dresden, Germany}
\date{\today}
\begin{abstract}
The nonconservative elastic responses of active solids have driven a recent explosion of interest in two-dimensional ``odd'' elasticity: small, linear deformations of these Cauchy elastic solids enable new behaviour absent from classical, passive elasticity. Here, we establish the description of large, nonlinear deformations of isotropic two-dimensional Cauchy elastic solids. We apply our framework to the Rivlin problem, perhaps the simplest problem of elasticity lacking a linear analogue: a square deforms under dead load tractions. Surprisingly, we find that oddness suppresses the bifurcations of a passive Rivlin square. By contrast, we discover that the bifurcations of a three-dimensional Rivlin cube survive oddness even though there is no isotropic, odd linear elasticity in three dimensions. Our results thus form the basis for describing large deformations of active, biological solids while revealing their unexpected nonlinear behaviour that arises even in minimal problems.
\end{abstract}

\maketitle
\end{CJK*}

The modern field theory of nonlinear elasticity is often based on the concept of hyperelasticity proposed by Green~\cite{ogden,truesdell}, in which the mechanical stresses are derived from a conserved strain energy function. Cauchy elasticity~\cite{ogden,yavari2025} is based instead on a direct, local relation between stresses and the deformation gradient tensor. The consequent lack of a conserved energy allows the net work performed by stress along a closed path to be nonzero in a Cauchy elastic material. For this reason, Cauchy elasticity received little attention until metamaterials and biological materials provided examples of systems with local energy sources and sinks that offer a physical realisation of this central feature of Cauchy elasticity.

Indeed, the advent of active matter has brought about considerable recent interest in the physics of ``odd'' elasticity \mbox{\cite{scheibner_odd,annualreview_odd,chen21,kole21,banerjee21,bordiga2022,tan2022,lier22,surowka23,shankar24,zhang24,lin24,walden25,lee25,zhou25,chao26,bo26,Mou26,lee26,Binysh26}}: linear Cauchy elasticity by another name would smell as sweet~\cite{* [] [{; for a modern edition, see, e.g., }] shakespeare,*rj,yavari2025}. Two-dimensional (2D) linear odd elasticity adds the additional elastic moduli allowed by removing the symmetries implied by conservation of energy and angular moment to the stress-strain relation of passive linear elasticity. These odd moduli couple, for example, compression to rotation, and hence give rise to novel behaviour that is not allowed in passive elastic solids. Interestingly, this linear phenomenology is limited to 2D: there is no three-dimensional (3D) isotropic linear odd elasticity~\cite{avron98,scheibner_odd,annualreview_odd}.

This treatment of 2D odd elasticity has however been limited to the linear regime and hence to small deformations. This cannot describe the large deformations of active biological solids.

Here, we set up a constitutive framework for finite deformations of 2D odd elastic solids. We apply our framework to the classical Rivlin problem~\cite{rivlin_cube,ball1983}, the bifurcation of an elastic square under dead load tractions that is absent from linear elasticity. Strikingly, we find that the rich behaviour of a passive Rivlin square does not appear for an odd Rivlin square, which only allows trivial deformations in general. We then construct a nonlinear theory of isotropic incompressible odd elasticity in 3D. Although there is no linear isotropic odd elasticity in 3D, we discover non-trivial behaviour in an odd Rivlin cube.

\paragraph{2D isotropic nonlinear odd elasticity.} We consider a 2D deformation that maps each point $\vec{X}\in\mathbb{R}^2$ in a body to its deformed position $\vec{x}(\vec{X})\in\mathbb{R}^2$. The deformation gradient tensor associated to this map is $\tens{F}=\partial\vec{x}/\partial\vec{X}$. Cauchy elasticity postulates the relation between the Cauchy stress $\tens{T}$ and the deformation. For a simple material, assumed here, this relation is instantaneous and local, viz., $\tens{T}=\tens{T}(\tens{F})$.

This relation must satisfy the fundamental requirement of \emph{objectivity}, i.e., that the physics do not change under rigid-body motions~\cite{ogden,truesdell,goriely,yavari2025}. Mathematically, this is expressed by the condition $\tens{T}(\tens{QF})=\tens{QT}(\tens{F})\tens{Q}^\top$ for all deformations $\tens{F}$ and all rotations $\tens{Q}\in\text{SO}(2)$~\cite{truesdell}.

Moreover, this relation must obey the symmetries of the material. For an isotropic material, this condition reads $\tens{Q}\tens{T}(\tens{F}) \tens{Q}^\top = \tens{T}\bigl(\tens{Q}\tens{F}\tens{Q}^\top\bigr)$ for all deformations $\tens{F}$ and all rotations $\tens{Q}\in\text{SO}(2)$~\cite{truesdell}.

Combining these conditions, ${\tens{T}(\tens{QF})\!=\!\tens{T}\bigl((\tens{QF})\tens{Q}^\top\bigr)}$ for all $\tens{QF}$ and all $\tens{Q}^\top\in\text{SO}(2)$; equivalently, ${\tens{T}(\tens{F})=\tens{T}(\tens{FQ})}$ for all $\tens{F}$ and all $\tens{Q}\in\text{SO}(2)$. We now introduce the polar decomposition $\tens{F}=\tens{VR}$, where $\tens{V}=\tens{V}^\top$ and $\tens{R}\in\text{SO}(2)$. Then, choosing $\tens{Q}=\tens{R}^\top$ in the last condition, we obtain $\tens{T}(\tens{F})=\tens{T}(\tens{FQ})=\tens{T}\bigl((\tens{VR})\tens{R}^\top\bigr)=\tens{T}(\tens{V})$. Moreover, we have $\tens{T}\bigl(\tens{Q}_\mathbfsf{1}\tens{F}\tens{Q}_\mathbfsf{2}^\top\bigr)=\tens{Q}_\mathbfsf{1}\tens{T}\bigl(\tens{F}\tens{Q}_\mathbfsf{2}^\top\bigr)\tens{Q}_\mathbfsf{1}^\top=\tens{Q}_\mathbfsf{1}\tens{T}(\tens{F})\tens{Q}_\mathbfsf{1}^\top$ for all $\tens{Q}_\mathbfsf{1},\tens{Q}_\mathbfsf{2}\in\text{SO}(2)$. Taking $\tens{Q}_\mathbfsf{2}=\tens{Q}_\mathbfsf{1}\tens{R}$, and using the previous result, this becomes ${\tens{T}\bigl(\tens{QVQ}^\top\bigr)=\tens{Q}\tens{T}(\tens{V})\tens{Q}^\top}$ for all $\tens{Q}\in\text{SO}(2)$. Now let $\smash{\tens{B}=\tens{FF}^\top=\tens{V}^2}$ be the left Cauchy--Green tensor~\cite{ogden,truesdell,goriely}. Write ${\tens{T}(\tens{V})\!=\!\mathcalbf{T}\smash{\bigl(\tens{V}^2\bigr)}\!=\!\mathcalbf{T}(\tens{B})}$, so\linebreak ${\tens{Q}\mathcalbf{T}(\tens{B})\tens{Q}^\top=\tens{T}\bigl(\smash{\tens{QVQ}}^\top\bigr)=\mathcalbf{T}\bigl((\smash{\tens{QVQ}}^\top)^2\bigr)=\mathcalbf{T}\bigl(\smash{\tens{QBQ}}^\top\bigr)}$ for all $\tens{Q}\in\text{SO}(2)$. This condition says that $\tens{T}$ is an isotropic function of $\tens{B}$~\footnote{In hyperelasticity~\cite{goriely}, an analogous argument establishes that the energy density $W$ is a (scalar) isotropic function of $\tens{B}$, but we are not aware of a reference for this result in Cauchy elasticity.}.

The irreducible representations of 2D isotropic second-rank tensors are known in general~\cite{zheng94}. These results now imply that
\begin{subequations}\label{eq:T}
\begin{align}
\tens{T}=t_1\tens{I}+t_2\tens{B}+t_3[\tens{B},\teps]+t_4\teps,\label{eq:T1}
\end{align}
in which $\tens{I}$ is the identity, $\teps$ is the 2D alternating symbol, and $[\tens{B},\teps]=\tens{B}\teps-\teps\tens{B}$, and where $t_1$, $t_2$, $t_3$, $t_4$ are isotropic scalar functions, so~\cite{zheng94} are functions of $\mathcal{I}=\tr{\tens{B}}$ and $\mathcal{J}=\det{\tens{F}}$, since $\det{\tens{B}}=\mathcal{J}^2$.

Geometrically, $\teps$ is rotation by $\pi/2$. This term is allowed by objectivity and isotropy to appear in 2D Cauchy elasticity because rotations commute in 2D. They do not commute in 3D however; fundamentally, this is why there is no linear isotropic odd elasticity in 3D~\cite{avron98}.

By contrast $\tens{T} = t_1\tens{I}+t_2\tens{B}$  in 2D hyperelasticity~\footnote{The corresponding result in 3D hyperelasticity is well-known~\cite{goriely,rivlin97}, but we are not aware of an exact reference for the 2D result: since the energy density $W$ is a scalar function of $\tens{B}$ only~\cite{Note1}, it is a function of $\mathcal{I},\mathcal{J}$ only; the definition $\tens{T}=\mathcal{J}^{-1}\tens{F}\partial W/\partial\tens{F}$~\cite{goriely} then yields the representation $\tens{T}=t_1\tens{I}+t_2\tens{B}$, where $t_1=\partial W/\partial\mathcal{J}$ and $t_2=2\mathcal{J}^{-1}\partial W/\partial\mathcal{I}$. Thus, unlike in Eq.~\eqref{eq:T1}, $t_1,t_2$ are not arbitrary. This means that $2\partial t_1/\partial\mathcal{I} \neq t_2+\mathcal{J}\partial t_2/\partial\mathcal{J}$ introduces odd behaviour, too. We will raise a similar point in our discussion of 3D nonlinear odd elasticity.}.\nocite{rivlin97} Thus the last two terms in Eq.~\eqref{eq:T1} introduce odd behaviour. The second of these additionally breaks angular momentum conservation, because $\tens{T}\neq\tens{T}^\top$ if $t_4\neq 0$.

It is useful to express Eq.~\eqref{eq:T1} in terms of the deviatoric part of $\tens{B}$, $\dev{\tens{B}}=\tens{B}-(\mathcal{I}/2)\tens{I}$. Renaming coefficients, we therefore write
\begin{align}
\tens{T}=t_1\tens{I}+t_2\dev{\tens{B}}+t_3[\tens{B},\teps]+t_4\teps.\label{eq:T2}
\end{align}
\end{subequations}
We also require that there be no residual stresses, i.e., $\tens{T}=\mathbfsf{0}$ if $\tens{F}=\tens{I}$. This implies that $t_1\tens{I}+t_4\teps=\mathbfsf{0}$ and hence $t_1=t_4=0$ if $\mathcal{I}=2$, $\mathcal{J}=1$.

To understand the physical meaning of $t_1,t_2,t_3,t_4$, we linearise Eq.~\eqref{eq:T2} around $\tens{F}=\tens{I} \Rightarrow \mathcal{I} =2,\mathcal{J}=1$ by introducing a small displacement $\vec{x}=\vec{X}+\vec{u}$, so that $\tens{F}=\tens{I}+\vec{\nabla u}$ with $\|\vec{\nabla u}\|\ll 1$. In the Supplemental Material~\footnote{See Supplemental Material at [url to be inserted], which includes Refs.~\cite{surowka24,Chen2017,Zhao2021,ogden,goriely,Overvelde2016,yavari2025}, for details of the calculations.}, we show that, in this limit, \nocite{surowka24,Chen2017,Zhao2021,ogden,goriely,Overvelde2016,yavari2025}
\begin{subequations}
\begin{align}
\tens{T}= \tens{t}+\mathcal{O}\bigl(\|\vec{\nabla u}\|^2\bigr),
\end{align}
with
\begin{align}
\tens{t}&=t_1^{(1)}(\vec{\nabla\cdot\vec{u}})\tens{I}+t_2^{(0)}\dev{\tens{S}}+t_3^{(0)}[\tens{S},\teps]\nonumber\\
&\qquad+t_4^{(1)} (\vec{\nabla\cdot\vec{u}})\teps,\label{eq:Tlin}
\end{align}
\end{subequations}
where $\tens{S}=\bigl(\vec{\nabla u}+\vec{\nabla u}^{\top}\bigr)/2$, and $t_1^{(1)},t_2^{(0)},t_3^{(0)},t_4^{(1)} $ result from expansions of $t_1,t_2,t_3,t_4$ near $\mathcal{I}=2,\mathcal{J}=1$. We map this result onto the constitutive equations of linear two-dimensional odd elasticity~\cite{surowka24}, $\tens{t}=\tens{C}:\vec{\nabla u}$, with
\begin{align}
C_{ijk\ell} &= \kappa\,\delta_{ij}\delta_{k\ell}  + \mu\big(\delta_{ik}\delta_{j\ell} +\delta_{i\ell}\delta_{jk} - \delta_{ij}\delta_{k\ell}\big) \nonumber\\
&\qquad+A\,\varepsilon_{ij}\delta_{k\ell}-K\big(\varepsilon_{ik}\delta_{j\ell} + \varepsilon_{j\ell}\delta_{ik}\big),
\end{align}
in which $\kappa$ is the bulk modulus, $\mu$ is the shear modulus, and $A,K$ are odd moduli~\cite{scheibner_odd}. This determines $\smash{t_1^{(1)}}=\kappa$, $\smash{t_2^{(0)}}=2\mu$, $\smash{t_3^{(0)}}=-K$, $\smash{t_4^{\smash{(1)}}}=-A$~\cite{Note3}, and hence exhbits the physical meaning of $t_1,t_2,t_3,t_4(\mathcal{I},\mathcal{J})$, but does not of course determine them uniquely.

In what follows, we specialise to the simplest odd constitutive relations: we require that $t_1,t_2,t_3,t_4$ be independent of $\mathcal{I}$ and choose their dependence on $\mathcal{J}$ by analogy with 2D compressible neo-Hookean elasticity~\cite{[{The constitutive relations of neo--Hookean elasticity are, again, better known in the 3D case~\cite{ogden,truesdell,goriely}; for the 2D case, see, e.g., }] [{ and }] Chen2017,*Zhao2021}. With this, we reduce Eqs.~\eqref{eq:T} to
\begin{subequations}
\begin{align}
\tens{T}=\kappa(\mathcal{J}-1)\tens{I}+\dfrac{\mu}{\mathcal{J}^2}\dev{\tens{B}}-\dfrac{K}{2\mathcal{J}^2}[\tens{B},\teps]-A(\mathcal{J}-1)\teps.\label{eq:TNH}
\end{align}
In particular, $\tens{T}=\kappa(\mathcal{J}-1)\tens{I}+\mu\mathcal{J}^{-2}\dev{\tens{B}}$ if ${A=K=0}$. This is the  stress in a neo-Hookean solid with strain energy ${W = \bigl[\mu\bigl(\mathcal{J}^{-1}\mathcal{I} - 2\bigr) + {\kappa}(\mathcal{J}-1)^{2}\bigr]/2}$ \cite{Chen2017,vnote}.

We extend this relation to the incompressible case $\mathcal{J}=1$ by introducing the pressure $p$. On absorbing the isotropic part of $\tens{B}$ into $p$,
\begin{align}
\tens{T}=-p\tens{I}+\mu\tens{B}-\dfrac{K}{2}[\tens{B},\teps].
\end{align}
\end{subequations}
In this model, ${\tens{T}=\tens{T}^\top}$, so angular momentum is conserved, but this need not be the case. Indeed, for linearised deformations, $\mathcal{J} = 1\Rightarrow\vec{\nabla\cdot u}=0$, so $\tens{t}=\tens{t}^\top$ from Eq.~\eqref{eq:Tlin} and angular momentum is conserved. However, $\mathcal{J}=1$, $\mathcal{I}\neq2$ and hence $t_4\neq 0$ are allowed for finite deformations. Thus angular momentum conservation can be broken in isotropic incompressible odd elasticity for finite deformations only.

\begin{figure}[b]
\centering
\includegraphics[width=\linewidth]{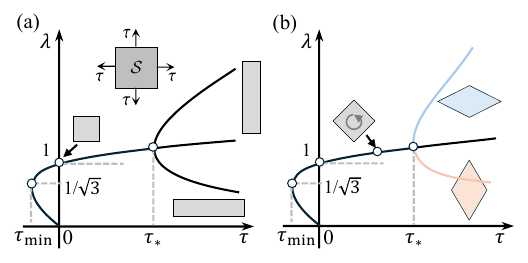}
\caption{The compressible 2D Rivlin problem. (a)~Bifurcation diagram, for $A=K=0$, of a square $\mathcal{S}$ under dead load tractions~$\tau$ (inset): For ${\tau<0}$ (compression), the square shrinks past $\tau=\tau_{\text{min}}$ towards the fully compressed state $\lambda = \tau =0$. For $\tau>0$ (traction), the square expands. At $\tau=\tau_\ast$, the square branch bifurcates to two rectangular branches. (b)~Corresponding bifurcation diagram for an odd square satisfying the parameter matching condition $\mu A + \kappa K= 0$. The square rotates for $\tau\neq 0$. At $\tau= \tau_\ast$, the rotated square bifurcates to a parallelogram.}
\label{fig1}
\end{figure}

\paragraph{The odd compressible Rivlin square.} We apply this minimal model of 2D isotropic nonlinear odd elasticity to the Rivlin problem \cite{rivlin_cube,ball1983,rivlin2003,haughton2005,tarantino2008,Overvelde2016,Mihai2019}: a square~$\mathcal{S}$ deforms under dead load tractions $\tau$ applied to its undeformed faces~\footnote{The Rivlin problem is a solvable paradigm of nonlinear elasticity~\cite{Mihai2019}, but experimental realisations are tricky, if not impossible~\cite{Overvelde2016} because, with the assumed dead loads, the total force on each side of $\mathcal{S}$ is constant, so the (Cauchy) stress on the deformed faces is not constant.}. At a critical load $\tau=\tau_\ast$, a bifurcation breaks the symmetry of the square shape~\figref{fig1}{a}. This bifurcation is absent from linear elasticity. The dead load boundary conditions are ${\tens{P}\cdot\vec{N}=\tau\vec{N}}$ on the boundary of~$\mathcal{S}$, where $\tens{P}=\mathcal{J}\tens{TF}^{-\top}$ is the first Piola--Kirchhoff stress \cite{goriely} and $\vec{N}$ is the (outward) normal to $\mathcal{S}$. For homogeneous deformations, this implies ${\tens{P}=\tau\tens{I}}$.

We start by solving for isotropic deformations $\tens{F}=\lambda\tens{R}$, where $\lambda>0$ is an isotropic stretch and $\tens{R}$ is rotation by an angle $\theta$. In the Supplemental Material~\cite{Note3}, we obtain
\begin{align}
&\tau=\bigl(k^2+A^2\bigr)^{1/2}\lambda\bigl(\lambda^2-1\bigr),&&\tan{\theta}=\dfrac{A}{\kappa}.
\end{align}
Thus the square shrinks ($\lambda<1$) as the compression ${\tau<0}$ increases in magnitude, until a minimum value $\tau=\tau_{\min}$ \figref{fig1}{a}. Because the dead load tractions apply a constant force per undeformed area, the compressive force becomes smaller as the square shrinks towards the fully compressed $\lambda=0$.

Next, we seek anisotropic homogeneous deformations: we introduce the polar decomposition $\tens{F}=\tens{VR}$, with ${\tens{V}=\operatorname{diag}(\lambda_1,\lambda_2)}$. We find~\cite{Note3} that an anisotropic solution, $\lambda_1\neq\lambda_2$ exists only if $\mu A+\kappa K=0$. This is satisfied for an even square ($A=K=0$). In this case, $\tens{R}=\tens{I}$ and these solutions define two rectangular branches bifurcating off the square branch at $\tau=\tau_\ast$ \figref{fig1}{a}, corresponding to $\lambda=\lambda_\ast$, with~\cite{Note3}
\begin{align}
\lambda_\ast^2=\dfrac{1+\sqrt{1+8\mu/\kappa}}{2}.
\end{align}
The odd case is more interesting: $\mu A+\kappa K\neq 0$ in general, so there is no anisotropic homogeneous deformation and the square does not bifurcate. This result for the nonlinear Rivlin problem contrasts with problems of linear elasticity, in which oddness gives rise to behaviour absent from even materials \cite{scheibner_odd,annualreview_odd,chen21,kole21,banerjee21,bordiga2022,tan2022,lier22,surowka23,shankar24,zhang24,lin24,walden25,lee25,zhou25,chao26,Mou26,bo26,lee26,Binysh26}. If however ${\mu A+\kappa K=0}$, then $\tens{R}\neq\tens{I}$, and the rotated square bifurcates to a parallelogram at the same $\tau=\tau_\ast$, $\lambda=\lambda_\ast$ \figref{fig1}{b}.

It is known that the even Rivlin problem only admits homogeneous deformations under the mild assumptions of Ref.~\cite{ball1983}. This argument does not carry over to the odd case, so we analyse inhomogeneous deformations by solving the equations of incremental elasticity~\cite{ogden}, around the isotropic deformation $\tens{F}=\lambda\tens{R}$, for a finite-wavelength increment $\delta\vec{u}=\vec{a}\,\e^{\i \vec{q}\cdot\vec{X}}$, in which $\vec{q}$ is a polarisation vector, and $\vec{a}$ is the magnitude of the increment. This calculation, which we postpone to the Supplemental Material~\cite{Note3}, leads to $\tens{Q}\cdot\vec{a}=\vec{0}$, in which $\tens{Q}$ is interpreted as an acoustic tensor~\cite{ogden,Roychowdhury2025}. There is a nontrivial solution, $\vec{a}\neq\vec{0}$, if and only if~\cite{Note3}
\begin{align}
0=\det{\tens{Q}}=\bigl(\kappa\mu-AK\bigr)+\dfrac{\mu^2+K^2}{\lambda^4},
\end{align}
which requires $\kappa\mu-AK<0$. Hence inhomogeneous deformations first arise at $\tau=\tau_{\ast\ast}$, corresponding to 
\begin{align}
\lambda=\lambda_{\ast\ast}=\left(\dfrac{\mu^2+K^2}{AK-\kappa\mu}\right)^{1/4}.
\end{align}
Examples of these inhomogeneous deformation modes are shown in \textwholefigref{fig2}.

We expect odd materials to still resist compression or simple shear, i.e., $\kappa,\mu>0$~\cite{[{For even materials, thermodynamic stability, $W\geqslant 0$, requires $\kappa\geqslant 0$ and $\mu\geqslant 0$: see, e.g., }][{, but this argument cannot be made for odd materials.}] landaulifshitz}, so this shows that inhomogeneous solutions exist only if the material is ``sufficiently'' odd, $AK>\kappa\mu>0$. Again, this is different from linear odd elasticity, where new behaviour arises for any $A,K\neq 0$. We also note that the condition $AK>\kappa\mu>0$ is incompatible with $\mu A+\kappa K=0$.

\begin{figure}[t]
\centering
\includegraphics[width=0.85\linewidth]{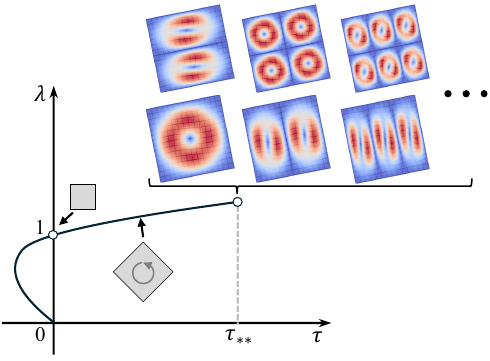}
\caption{Bifurcations of a very odd compressible Rivlin square. If $AK>\kappa\mu$, then inhomogeneous deformations bifurcate off the square branch at $\tau=\tau_{\ast\ast}$. Insets show possible modes~\cite{Note3}.}
\label{fig2}
\end{figure}

We can analyse the incompressible Rivlin square similarly. We find~\cite{Note3} the homogeneous bifurcation condition to be $(2\mu-\tau_\ast)^2+4K^2=0$. Thus an even square ($K=0$) bifurcates at $\tau_\ast=2\mu$, but, again, there is no bifurcation for an odd square ($K\neq 0$). Moreover, there are no inhomogeneous solutions in the incompressible case~\cite{Note3}.

\paragraph{3D isotropic, incompressible nonlinear odd elasticity.} In 3D, the Cauchy stress in an isotropic Cauchy elastic solid can be written as~\cite{yavari2025}
\begin{align}
 \tens{T}=-p\tens{I}+\alpha(\mathcal{I}_1,\mathcal{I}_2)\tens{B} + \beta(\mathcal{I}_1,\mathcal{I}_2)\tens{B}^{2},\label{eq:T3D}
\end{align}
in which $p$ is pressure, $\alpha,\beta$ are (arbitrary) functions of the principal invariants $\mathcal{I}_1=\tr{\tens{B}}$, ${\mathcal{I}_2=[(\tr{\tens{B}})^2-\tr{\tens{B}^2}]/2}$. For a hyperelastic material with isotropic energy density $W(\mathcal{I}_1,\mathcal{I}_2)$, the Cauchy stress is also of this form, with $\alpha=2W_{,\mathcal I_1}+2\mathcal I_1{W}_{,\mathcal I_2}$, $\beta=-2{W}_{,\mathcal I_2}$~\cite{goriely}, where commata denote differentiation. This implies ${W_{,\mathcal{I}_1}=(\alpha+\mathcal{I}_1\beta)/2}$, $W_{,\mathcal{I}_2}=-\beta/2$. A necessary and sufficient condition for such $W$ to exist is $(W_{,\mathcal{I}_1})_{,\mathcal{I}_2}=(W_{,\mathcal{I}_2})_{,\mathcal{I}_1}$ \cite{boas2006}, which reduces to
\begin{align}
\frac{\partial \alpha}{\partial \mathcal I_2}+ \mathcal I_1\,\frac{\partial \beta}{\partial \mathcal I_2}+\frac{\partial \beta}{\partial \mathcal I_1}=0.
\label{eq:integrability}
\end{align}
This is not however a sufficient condition for the material described by Eq.~\eqref{eq:T3D} to be even, because of the additional constraint of thermodynamic stability~\cite{landaulifshitz}, that $W\geqslant 0$ for all deformations. This makes the point that even a hyperelastic solid can be odd.

By analogy with polynomial hyperelastic constitutive laws \cite{rivlin1951,ogden}, we can expand $\alpha,\beta$ as power series in ${\mathcal{I}_1-3}$, $\mathcal{I}_2-3$~\cite{Note3}. In what follows, we focus on the minimal case in which this expansion is truncated at linear order, 
\begin{subequations}\label{eq:ab}
\begin{align}
\alpha(\mathcal{I}_1,\mathcal{I}_2)&=\alpha_{00}+\alpha_{10}(\mathcal{I}_1-3)+\alpha_{01}(\mathcal{I}_2-3),\\
\beta(\mathcal{I}_1,\mathcal{I}_2)&=\beta_{00}+\beta_{10}(\mathcal{I}_1-3)+\beta_{01}(\mathcal{I}_2-3).
\end{align}
\end{subequations}
Clearly, condition \eqref{eq:integrability} holds for all $\mathcal{I}_1,\mathcal{I}_2$ if and only if $\alpha_{01}=-\beta_{10}$ and $\beta_{01}=0$. Expansions~\eqref{eq:ab} thus define a six-dimensional space of Cauchy elastic materials, with a four-dimensional subspace of hyperelastic materials.

\paragraph{The odd Rivlin cube.} We apply this minimal model of 3D incompressible odd elasticity to the Rivlin cube \cite{rivlin_cube,ball1983,rivlin2003,haughton2005,tarantino2008,Mihai2019}. We seek a homogeneous deformation ${\tens{F}=\mathrm{diag}(\lambda_1,\lambda_2,\lambda_3)}$, with ${\lambda_1\lambda_2\lambda_3=1}$ by incompressibility. Then $\tens{B}=\mathrm{diag}(\lambda_1^2,\lambda_2^2,\lambda_3^2)$, ${\tens{B}^2=\mathrm{diag}(\lambda_1^4,\lambda_2^4,\lambda_3^4)}$. On substitution into Eq.~\eqref{eq:T3D}, this determines the first Piola--Kirchhoff stress $\tens{P}=\tens{TF}^{-\top}$. For dead loads $\tau$ and the assumed homogeneous deformations, $\tens{P}=\tau\tens{I}$. Hence
\begin{align}
\tau\lambda_i=-p+\alpha\lambda_i^2+\beta\lambda_i^4,\quad i=1,2,3.
\end{align}
Subtracting these equations from each other yields
\begin{align}
(\lambda_i-\lambda_j)(m_{i,j}-\tau)=0,\quad i,j=1,2,3,\label{eq:rivlin2}
\end{align}
where $m_{i,j}=(\lambda_i +\lambda_j)\left[\alpha + \beta\left(\lambda_i^2 + \lambda_j^2\right)\right]$. This allows the three solution classes of the Rivlin problem~\cite{rivlin_cube,ball1983,Mihai2019}:
\begin{enumerate}[label=(\arabic{enumi}),widest=2,leftmargin=*,itemsep=0pt,topsep=3pt]
\item The trivial solution is $\lambda_1=\lambda_2=\lambda_3=1$.
\item In a biaxial deformation, without loss of generality, $\lambda_1=\lambda_2=\lambda$, ${\lambda_3=\lambda^{-2}}$, with $\lambda\neq1$. Equations~\eqref{eq:rivlin2} then reduce to $\tau=m_{1,2}$, whence
\begin{align}
\tau(\lambda) = \frac{\alpha\big(\lambda^4-\lambda^{-2}\big) +\beta\big(\lambda^6-\lambda^{-6}\big)}{\lambda^3-1}. \label{eq:taubiaxial}
\end{align}
To determine the  point $\tau=\tau_\ast$ at which this biaxial solution bifurcates off trivial one, we take the limit $\lambda\to 1$ in Eq.~\eqref{eq:taubiaxial}. This yields $\tau\to\tau_\ast=2\mu$.
\item In a triaxial deformation, $\lambda_1\neq\lambda_2\neq\lambda_3\neq\lambda_1$, so $m_{1,2}=m_{2,3}=m_{3,1}=\tau$ from Eq.~\eqref{eq:rivlin2}. This implies that $
0=m_{1,2}-m_{3,1}=(\lambda_2-\lambda_3)(\alpha+\beta r)$, where $r=\lambda_1^2+\lambda_2^2+\lambda_3^2+\lambda_1\lambda_2+\lambda_2\lambda_3+\lambda_3\lambda_1$. Since $\lambda_2\neq\lambda_3$,
\begin{align}
\alpha+\beta r=0.\label{eq:triaxialcond}
\end{align}
Conversely, if $\alpha+\beta r=0$, then $m_{1,2}=m_{2,3}=m_{3,1}$. Now, on the trivial branch, $m_{1,2}=m_{2,3}=m_{3,1}=2\mu$ and $r=6$, so $\alpha_{00}+6\beta_{00} =0$ is a necessary condition for a triaxial solution to bifurcate off the trivial branch, which happens at $\tau=\tau_\ast=2\mu$, too. In the Supplemental Material~\cite{Note3}, we solve Eq.~\eqref{eq:triaxialcond} by expansion near $\lambda_1=\lambda_2=\lambda_3=1$ to obtain necessary and sufficient conditions for such a triaxial bifurcation off the trivial branch, $\alpha_{00}+6\beta_{00} =0$ and $4(\alpha_{01}+\alpha_{10})+5\beta_{00}+24(\beta_{01}+\beta_{10})=0$.
\end{enumerate}
The classical biaxial bifurcation of an incompressible Rivlin cube \mbox{\cite{rivlin_cube,ball1983}} thus persists in the space of Cauchy elastic materials defined by Eqs.~\eqref{eq:ab}. There is a subset of these materials with an additional, triaxial bifurcation of the trivial branch, but the hyperelastic space in which this additional bifurcation happens is only a two-dimensional subspace of this four-dimensional space. In other words, this additional bifurcation behaviour of the Rivlin cube requires more parameter matching conditions for an even material than for an odd one, in contrast with our results for the Rivlin square.

In this Letter, we have constructed minimal models of nonlinear isotropic odd elasticity and, through the minimal example of the Rivlin problem, shown the surprising behaviour they afford: In 2D, oddness suppresses, in general, the behaviour of a passive Rivlin square, in contrast to the rich behaviour of isotropic linear odd elasticity in 2D. Somewhat conversely, there is no isotropic linear odd elasticity in 3D, but the bifurcation behaviour of a Rivlin cube survives oddness, and new bifurcations can appear. More conceptually, our work has taken some of the first steps in the programme of nonlinear Cauchy elasticity set out in Ref.~\cite{yavari2025}.

We have analysed one possible, minimal extension of linear odd elasticity, which is of course not unique. For example, we have focused here on materials with a positive linear shear modulus, $\mu>0$, but purely nonlinear elasticity~\cite{zhao2025} at $\mu=0$, with a positive nonlinear shear modulus is allowed thermodynamically. A fuller understanding of the effect of these different nonlinearities on different mechanical bifurcations therefore remains an important question for future work.

Moreover, we have analysed isotropic materials. Anisotropy can lead to linear odd moduli even in 3D~\cite{scheibner_odd}. What additional terms does microscopic structure, resulting, e.g., from nematic order, allow in the nonlinear relation between $\tens{T}$ and $\tens{F}$? These additional terms are well-studied in hyperelasticity~\cite{goriely}, but we are not aware of corresponding results for Cauchy elasticity. Classifying the tensor functions of $\tens{F}$ satisfying objectivity and the symmetries allowed by this additional structure therefore remains a challenge for future work.

More abstractly, we have introduced a natural polynomial expansion of the Cauchy stress $\tens{T}$ in ${\mathcal{I}_1-3}$, $\mathcal{I}_2-3$ in Eq.~\eqref{eq:T3D}, motivated by the analogous expansion of the energy density $W$ that defines polynomial hyperelastic laws~\cite{ogden,rivlin1951}. However, even the lowest-order truncation of $\tens{T}$, with $\alpha(\mathcal{I}_1,\mathcal{I}_2)=\alpha_{00}$, $\beta(\mathcal{I}_1,\mathcal{I}_2)=\beta_{00}$, is seen to imply $W\propto2(\alpha_{00}+3\beta_{00})(\mathcal{I}_1-3)-2\beta_{00}(\mathcal{I}_2-3)+\beta_{00}(\mathcal{I}_1-3)^2$.\linebreak By contrast, the lowest-order polynomial (Mooney--Rivlin) law~\cite{goriely,rivlin1951}, ${W\propto w_{10}(\mathcal{I}_1-3)+w_{01}(\mathcal{I}_2-3)}$ is less nonlinear. In other words, these two natural expansions do not ``commute'', the consequences of which remain another open problem.

A more general question remains open, too: What are biological systems or metamaterials that realise the nonlinear odd effects that we have revealed here?

\begin{acknowledgments}
We thank Vincenzo Galgano for a conversation on tensor representations and gratefully acknowledge funding from the Max Planck Society.
\end{acknowledgments}

\bibliography{main.bib}

\begin{thebibliography}{7}%
\makeatletter
\providecommand \@ifxundefined [1]{%
 \@ifx{#1\undefined}
}%
\providecommand \@ifnum [1]{%
 \ifnum #1\expandafter \@firstoftwo
 \else \expandafter \@secondoftwo
 \fi
}%
\providecommand \@ifx [1]{%
 \ifx #1\expandafter \@firstoftwo
 \else \expandafter \@secondoftwo
 \fi
}%
\providecommand \natexlab [1]{#1}%
\providecommand \enquote  [1]{``#1''}%
\providecommand \bibnamefont  [1]{#1}%
\providecommand \bibfnamefont [1]{#1}%
\providecommand \citenamefont [1]{#1}%
\providecommand \href@noop [0]{\@secondoftwo}%
\providecommand \href [0]{\begingroup \@sanitize@url \@href}%
\providecommand \@href[1]{\@@startlink{#1}\@@href}%
\providecommand \@@href[1]{\endgroup#1\@@endlink}%
\providecommand \@sanitize@url [0]{\catcode `\\12\catcode `\$12\catcode
  `\&12\catcode `\#12\catcode `\^12\catcode `\_12\catcode `\%12\relax}%
\providecommand \@@startlink[1]{}%
\providecommand \@@endlink[0]{}%
\providecommand \url  [0]{\begingroup\@sanitize@url \@url }%
\providecommand \@url [1]{\endgroup\@href {#1}{\urlprefix }}%
\providecommand \urlprefix  [0]{URL }%
\providecommand \Eprint [0]{\href }%
\providecommand \doibase [0]{https://doi.org/}%
\providecommand \selectlanguage [0]{\@gobble}%
\providecommand \bibinfo  [0]{\@secondoftwo}%
\providecommand \bibfield  [0]{\@secondoftwo}%
\providecommand \translation [1]{[#1]}%
\providecommand \BibitemOpen [0]{}%
\providecommand \bibitemStop [0]{}%
\providecommand \bibitemNoStop [0]{.\EOS\space}%
\providecommand \EOS [0]{\spacefactor3000\relax}%
\providecommand \BibitemShut  [1]{\csname bibitem#1\endcsname}%
\let\auto@bib@innerbib\@empty
\bibitem [{\citenamefont {Ostoja-Starzewski}\ and\ \citenamefont
  {Sur\'owka}(2024)}]{surowka24}%
  \BibitemOpen
  \bibfield  {author} {\bibinfo {author} {\bibfnamefont {M.}~\bibnamefont
  {Ostoja-Starzewski}}\ and\ \bibinfo {author} {\bibfnamefont {P.}~\bibnamefont
  {Sur\'owka}},\ }\bibfield  {title} {\bibinfo {title} {Generalizing odd
  elasticity theory to odd thermoelasticity for planar materials},\ }\href
  {https://doi.org/10.1103/PhysRevB.109.064107} {\bibfield  {journal} {\bibinfo
   {journal} {Phys. Rev. B}\ }\textbf {\bibinfo {volume} {109}},\ \bibinfo
  {pages} {064107} (\bibinfo {year} {2024})}\BibitemShut {NoStop}%
\bibitem [{\citenamefont {Chen}\ \emph {et~al.}(2017)\citenamefont {Chen},
  \citenamefont {Chang},\ and\ \citenamefont {Qin}}]{Chen2017}%
  \BibitemOpen
  \bibfield  {author} {\bibinfo {author} {\bibfnamefont {L.}~\bibnamefont
  {Chen}}, \bibinfo {author} {\bibfnamefont {Z.}~\bibnamefont {Chang}},\ and\
  \bibinfo {author} {\bibfnamefont {T.}~\bibnamefont {Qin}},\ }\bibfield
  {title} {\bibinfo {title} {Elastic wave propagation in simple-sheared
  hyperelastic materials with different constitutive models},\ }\href
  {https://doi.org/https://doi.org/10.1016/j.ijsolstr.2017.07.027} {\bibfield
  {journal} {\bibinfo  {journal} {Int. J. Solids Struct.}\ }\textbf {\bibinfo
  {volume} {126--127}},\ \bibinfo {pages} {1} (\bibinfo {year}
  {2017})}\BibitemShut {NoStop}%
\bibitem [{\citenamefont {Zhao}\ and\ \citenamefont {Chang}(2021)}]{Zhao2021}%
  \BibitemOpen
  \bibfield  {author} {\bibinfo {author} {\bibfnamefont {S.}~\bibnamefont
  {Zhao}}\ and\ \bibinfo {author} {\bibfnamefont {Z.}~\bibnamefont {Chang}},\
  }\bibfield  {title} {\bibinfo {title} {Elastic wave velocities in finitely
  pre-stretched soft fibers},\ }\href
  {https://doi.org/https://doi.org/10.1016/j.ijsolstr.2021.111208} {\bibfield
  {journal} {\bibinfo  {journal} {Int. J. Solids Struct.}\ }\textbf {\bibinfo
  {volume} {233}},\ \bibinfo {pages} {111208} (\bibinfo {year}
  {2021})}\BibitemShut {NoStop}%
\bibitem [{\citenamefont {Ogden}(1997)}]{ogden}%
  \BibitemOpen
  \bibfield  {author} {\bibinfo {author} {\bibfnamefont {R.~W.}\ \bibnamefont
  {Ogden}},\ }\href@noop {} {\emph {\bibinfo {title} {Non-linear elastic
  deformations}}}\ (\bibinfo  {publisher} {Dover},\ \bibinfo {address}
  {Mineola, NY},\ \bibinfo {year} {1997})\ Chap.\ \bibinfo {chapter} {3 \& 4},
  pp.\ \bibinfo {pages} {140--222}\BibitemShut {NoStop}%
\bibitem [{\citenamefont {Goriely}(2017)}]{goriely}%
  \BibitemOpen
  \bibfield  {author} {\bibinfo {author} {\bibfnamefont {A.}~\bibnamefont
  {Goriely}},\ }\href@noop {} {\emph {\bibinfo {title} {The Mathematics and
  Mechanics of Biological Growth}}}\ (\bibinfo  {publisher} {Springer},\
  \bibinfo {address} {Berlin, Germany},\ \bibinfo {year} {2017})\
  Chap.~\bibinfo {chapter} {11}, pp.\ \bibinfo {pages} {261--344}\BibitemShut
  {NoStop}%
\bibitem [{\citenamefont {Overvelde}\ \emph {et~al.}(2016)\citenamefont
  {Overvelde}, \citenamefont {Dykstra}, \citenamefont {de~Rooij}, \citenamefont
  {Weaver},\ and\ \citenamefont {Bertoldi}}]{Overvelde2016}%
  \BibitemOpen
  \bibfield  {author} {\bibinfo {author} {\bibfnamefont {J.~T.~B.}\
  \bibnamefont {Overvelde}}, \bibinfo {author} {\bibfnamefont {D.~M.~J.}\
  \bibnamefont {Dykstra}}, \bibinfo {author} {\bibfnamefont {R.}~\bibnamefont
  {de~Rooij}}, \bibinfo {author} {\bibfnamefont {J.}~\bibnamefont {Weaver}},\
  and\ \bibinfo {author} {\bibfnamefont {K.}~\bibnamefont {Bertoldi}},\
  }\bibfield  {title} {\bibinfo {title} {Tensile instability in a thick elastic
  body},\ }\href {https://doi.org/10.1103/PhysRevLett.117.094301} {\bibfield
  {journal} {\bibinfo  {journal} {Phys. Rev. Lett.}\ }\textbf {\bibinfo
  {volume} {117}},\ \bibinfo {pages} {094301} (\bibinfo {year}
  {2016})}\BibitemShut {NoStop}%
\bibitem [{\citenamefont {Yavari}\ and\ \citenamefont
  {Goriely}(2025)}]{yavari2025}%
  \BibitemOpen
  \bibfield  {author} {\bibinfo {author} {\bibfnamefont {A.}~\bibnamefont
  {Yavari}}\ and\ \bibinfo {author} {\bibfnamefont {A.}~\bibnamefont
  {Goriely}},\ }\bibfield  {title} {\bibinfo {title} {Nonlinear {Cauchy}
  elasticity},\ }\href {https://doi.org/10.1007/s00205-025-02120-0} {\bibfield
  {journal} {\bibinfo  {journal} {Arch. Rational Mech. Anal.}\ }\textbf
  {\bibinfo {volume} {249}},\ \bibinfo {pages} {57} (\bibinfo {year}
  {2025})}\BibitemShut {NoStop}%
\end{thebibliography}%


\begin{thebibliography}{50}%
\makeatletter
\providecommand \@ifxundefined [1]{%
 \@ifx{#1\undefined}
}%
\providecommand \@ifnum [1]{%
 \ifnum #1\expandafter \@firstoftwo
 \else \expandafter \@secondoftwo
 \fi
}%
\providecommand \@ifx [1]{%
 \ifx #1\expandafter \@firstoftwo
 \else \expandafter \@secondoftwo
 \fi
}%
\providecommand \natexlab [1]{#1}%
\providecommand \enquote  [1]{``#1''}%
\providecommand \bibnamefont  [1]{#1}%
\providecommand \bibfnamefont [1]{#1}%
\providecommand \citenamefont [1]{#1}%
\providecommand \href@noop [0]{\@secondoftwo}%
\providecommand \href [0]{\begingroup \@sanitize@url \@href}%
\providecommand \@href[1]{\@@startlink{#1}\@@href}%
\providecommand \@@href[1]{\endgroup#1\@@endlink}%
\providecommand \@sanitize@url [0]{\catcode `\\12\catcode `\$12\catcode
  `\&12\catcode `\#12\catcode `\^12\catcode `\_12\catcode `\%12\relax}%
\providecommand \@@startlink[1]{}%
\providecommand \@@endlink[0]{}%
\providecommand \url  [0]{\begingroup\@sanitize@url \@url }%
\providecommand \@url [1]{\endgroup\@href {#1}{\urlprefix }}%
\providecommand \urlprefix  [0]{URL }%
\providecommand \Eprint [0]{\href }%
\providecommand \doibase [0]{https://doi.org/}%
\providecommand \selectlanguage [0]{\@gobble}%
\providecommand \bibinfo  [0]{\@secondoftwo}%
\providecommand \bibfield  [0]{\@secondoftwo}%
\providecommand \translation [1]{[#1]}%
\providecommand \BibitemOpen [0]{}%
\providecommand \bibitemStop [0]{}%
\providecommand \bibitemNoStop [0]{.\EOS\space}%
\providecommand \EOS [0]{\spacefactor3000\relax}%
\providecommand \BibitemShut  [1]{\csname bibitem#1\endcsname}%
\let\auto@bib@innerbib\@empty
\bibitem [{\citenamefont {Ogden}(1997)}]{ogden}%
  \BibitemOpen
  \bibfield  {author} {\bibinfo {author} {\bibfnamefont {R.~W.}\ \bibnamefont
  {Ogden}},\ }\href@noop {} {\emph {\bibinfo {title} {Non-linear elastic
  deformations}}}\ (\bibinfo  {publisher} {Dover},\ \bibinfo {address}
  {Mineola, NY},\ \bibinfo {year} {1997})\ Chap.\ \bibinfo {chapter} {3 \& 4},
  pp.\ \bibinfo {pages} {140--222}\BibitemShut {NoStop}%
\bibitem [{\citenamefont {Truesdell}\ and\ \citenamefont
  {Noll}(2004)}]{truesdell}%
  \BibitemOpen
  \bibfield  {author} {\bibinfo {author} {\bibfnamefont {C.}~\bibnamefont
  {Truesdell}}\ and\ \bibinfo {author} {\bibfnamefont {W.}~\bibnamefont
  {Noll}},\ }\href {https://doi.org/10.1007/978-3-662-10388-3_1} {\emph
  {\bibinfo {title} {The Non-Linear Field Theories of Mechanics}}},\ edited by\
  \bibinfo {editor} {\bibfnamefont {S.~S.}\ \bibnamefont {Antman}}\ (\bibinfo
  {publisher} {Springer},\ \bibinfo {address} {Berlin, Heidelberg},\ \bibinfo
  {year} {2004})\BibitemShut {NoStop}%
\bibitem [{\citenamefont {Yavari}\ and\ \citenamefont
  {Goriely}(2025)}]{yavari2025}%
  \BibitemOpen
  \bibfield  {author} {\bibinfo {author} {\bibfnamefont {A.}~\bibnamefont
  {Yavari}}\ and\ \bibinfo {author} {\bibfnamefont {A.}~\bibnamefont
  {Goriely}},\ }\bibfield  {title} {\bibinfo {title} {Nonlinear {Cauchy}
  elasticity},\ }\href {https://doi.org/10.1007/s00205-025-02120-0} {\bibfield
  {journal} {\bibinfo  {journal} {Arch. Rational Mech. Anal.}\ }\textbf
  {\bibinfo {volume} {249}},\ \bibinfo {pages} {57} (\bibinfo {year}
  {2025})}\BibitemShut {NoStop}%
\bibitem [{\citenamefont {Scheibner}\ \emph {et~al.}(2020)\citenamefont
  {Scheibner}, \citenamefont {Souslov}, \citenamefont {Banerjee}, \citenamefont
  {Sur{\'o}wka}, \citenamefont {Irvine},\ and\ \citenamefont
  {Vitelli}}]{scheibner_odd}%
  \BibitemOpen
  \bibfield  {author} {\bibinfo {author} {\bibfnamefont {C.}~\bibnamefont
  {Scheibner}}, \bibinfo {author} {\bibfnamefont {A.}~\bibnamefont {Souslov}},
  \bibinfo {author} {\bibfnamefont {D.}~\bibnamefont {Banerjee}}, \bibinfo
  {author} {\bibfnamefont {P.}~\bibnamefont {Sur{\'o}wka}}, \bibinfo {author}
  {\bibfnamefont {W.~T.}\ \bibnamefont {Irvine}},\ and\ \bibinfo {author}
  {\bibfnamefont {V.}~\bibnamefont {Vitelli}},\ }\bibfield  {title} {\bibinfo
  {title} {Odd elasticity},\ }\href
  {https://doi.org/doi.org/10.1038/s41567-020-0795-y} {\bibfield  {journal}
  {\bibinfo  {journal} {Nat. Phys.}\ }\textbf {\bibinfo {volume} {16}},\
  \bibinfo {pages} {475} (\bibinfo {year} {2020})}\BibitemShut {NoStop}%
\bibitem [{\citenamefont {Fruchart}\ \emph {et~al.}(2023)\citenamefont
  {Fruchart}, \citenamefont {Scheibner},\ and\ \citenamefont
  {Vitelli}}]{annualreview_odd}%
  \BibitemOpen
  \bibfield  {author} {\bibinfo {author} {\bibfnamefont {M.}~\bibnamefont
  {Fruchart}}, \bibinfo {author} {\bibfnamefont {C.}~\bibnamefont
  {Scheibner}},\ and\ \bibinfo {author} {\bibfnamefont {V.}~\bibnamefont
  {Vitelli}},\ }\bibfield  {title} {\bibinfo {title} {Odd viscosity and odd
  elasticity},\ }\href
  {https://doi.org/doi.org/10.1146/annurev-conmatphys-040821-125506} {\bibfield
   {journal} {\bibinfo  {journal} {Annu. Rev. Condens. Matter Phys.}\ }\textbf
  {\bibinfo {volume} {14}},\ \bibinfo {pages} {471} (\bibinfo {year}
  {2023})}\BibitemShut {NoStop}%
\bibitem [{\citenamefont {Chen}\ \emph {et~al.}(2021)\citenamefont {Chen},
  \citenamefont {Li}, \citenamefont {Scheibner}, \citenamefont {Vitelli},\ and\
  \citenamefont {Huang}}]{chen21}%
  \BibitemOpen
  \bibfield  {author} {\bibinfo {author} {\bibfnamefont {Y.}~\bibnamefont
  {Chen}}, \bibinfo {author} {\bibfnamefont {X.}~\bibnamefont {Li}}, \bibinfo
  {author} {\bibfnamefont {C.}~\bibnamefont {Scheibner}}, \bibinfo {author}
  {\bibfnamefont {V.}~\bibnamefont {Vitelli}},\ and\ \bibinfo {author}
  {\bibfnamefont {G.}~\bibnamefont {Huang}},\ }\bibfield  {title} {\bibinfo
  {title} {Realization of active metamaterials with odd micropolar
  elasticity},\ }\href {https://doi.org/0.1038/s41467-021-26034-z} {\bibfield
  {journal} {\bibinfo  {journal} {Nat. Commun.}\ }\textbf {\bibinfo {volume}
  {12}},\ \bibinfo {pages} {5935} (\bibinfo {year} {2021})}\BibitemShut
  {NoStop}%
\bibitem [{\citenamefont {Kole}\ \emph {et~al.}(2021)\citenamefont {Kole},
  \citenamefont {Alexander}, \citenamefont {Ramaswamy},\ and\ \citenamefont
  {Maitra}}]{kole21}%
  \BibitemOpen
  \bibfield  {author} {\bibinfo {author} {\bibfnamefont {S.~J.}\ \bibnamefont
  {Kole}}, \bibinfo {author} {\bibfnamefont {G.~P.}\ \bibnamefont {Alexander}},
  \bibinfo {author} {\bibfnamefont {S.}~\bibnamefont {Ramaswamy}},\ and\
  \bibinfo {author} {\bibfnamefont {A.}~\bibnamefont {Maitra}},\ }\bibfield
  {title} {\bibinfo {title} {Layered chiral active matter: Beyond odd
  elasticity},\ }\href {https://doi.org/10.1103/PhysRevLett.126.248001}
  {\bibfield  {journal} {\bibinfo  {journal} {Phys. Rev. Lett.}\ }\textbf
  {\bibinfo {volume} {126}},\ \bibinfo {pages} {248001} (\bibinfo {year}
  {2021})}\BibitemShut {NoStop}%
\bibitem [{\citenamefont {Banerjee}\ \emph {et~al.}(2021)\citenamefont
  {Banerjee}, \citenamefont {Vitelli}, \citenamefont {J\"ulicher},\ and\
  \citenamefont {Sur\'owka}}]{banerjee21}%
  \BibitemOpen
  \bibfield  {author} {\bibinfo {author} {\bibfnamefont {D.}~\bibnamefont
  {Banerjee}}, \bibinfo {author} {\bibfnamefont {V.}~\bibnamefont {Vitelli}},
  \bibinfo {author} {\bibfnamefont {F.}~\bibnamefont {J\"ulicher}},\ and\
  \bibinfo {author} {\bibfnamefont {P.}~\bibnamefont {Sur\'owka}},\ }\bibfield
  {title} {\bibinfo {title} {Active viscoelasticity of odd materials},\ }\href
  {https://doi.org/10.1103/PhysRevLett.126.138001} {\bibfield  {journal}
  {\bibinfo  {journal} {Phys. Rev. Lett.}\ }\textbf {\bibinfo {volume} {126}},\
  \bibinfo {pages} {138001} (\bibinfo {year} {2021})}\BibitemShut {NoStop}%
\bibitem [{\citenamefont {Bordiga}\ \emph {et~al.}(2022)\citenamefont
  {Bordiga}, \citenamefont {Piccolroaz},\ and\ \citenamefont
  {Bigoni}}]{bordiga2022}%
  \BibitemOpen
  \bibfield  {author} {\bibinfo {author} {\bibfnamefont {G.}~\bibnamefont
  {Bordiga}}, \bibinfo {author} {\bibfnamefont {A.}~\bibnamefont
  {Piccolroaz}},\ and\ \bibinfo {author} {\bibfnamefont {D.}~\bibnamefont
  {Bigoni}},\ }\bibfield  {title} {\bibinfo {title} {A way to hypo-elastic
  artificial materials without a strain potential and displaying flutter
  instability},\ }\href
  {https://doi.org/https://doi.org/10.1016/j.jmps.2021.104665} {\bibfield
  {journal} {\bibinfo  {journal} {J. Mech. Phys. Solids}\ }\textbf {\bibinfo
  {volume} {158}},\ \bibinfo {pages} {104665} (\bibinfo {year}
  {2022})}\BibitemShut {NoStop}%
\bibitem [{\citenamefont {Tan}\ \emph {et~al.}(2022)\citenamefont {Tan},
  \citenamefont {Mietke}, \citenamefont {Li}, \citenamefont {Chen},
  \citenamefont {Higinbotham}, \citenamefont {Foster}, \citenamefont {Gokhale},
  \citenamefont {Dunkel},\ and\ \citenamefont {Fakhri}}]{tan2022}%
  \BibitemOpen
  \bibfield  {author} {\bibinfo {author} {\bibfnamefont {T.~H.}\ \bibnamefont
  {Tan}}, \bibinfo {author} {\bibfnamefont {A.}~\bibnamefont {Mietke}},
  \bibinfo {author} {\bibfnamefont {J.}~\bibnamefont {Li}}, \bibinfo {author}
  {\bibfnamefont {Y.}~\bibnamefont {Chen}}, \bibinfo {author} {\bibfnamefont
  {H.}~\bibnamefont {Higinbotham}}, \bibinfo {author} {\bibfnamefont {P.~J.}\
  \bibnamefont {Foster}}, \bibinfo {author} {\bibfnamefont {S.}~\bibnamefont
  {Gokhale}}, \bibinfo {author} {\bibfnamefont {J.}~\bibnamefont {Dunkel}},\
  and\ \bibinfo {author} {\bibfnamefont {N.}~\bibnamefont {Fakhri}},\
  }\bibfield  {title} {\bibinfo {title} {Odd dynamics of living chiral
  crystals},\ }\href {https://doi.org/10.1038/s41586-022-04889-6} {\bibfield
  {journal} {\bibinfo  {journal} {Nature}\ }\textbf {\bibinfo {volume} {607}},\
  \bibinfo {pages} {287} (\bibinfo {year} {2022})}\BibitemShut {NoStop}%
\bibitem [{\citenamefont {Lier}\ \emph {et~al.}(2022)\citenamefont {Lier},
  \citenamefont {Armas}, \citenamefont {Bo}, \citenamefont {Duclut},
  \citenamefont {J\"ulicher},\ and\ \citenamefont {Sur\'owka}}]{lier22}%
  \BibitemOpen
  \bibfield  {author} {\bibinfo {author} {\bibfnamefont {R.}~\bibnamefont
  {Lier}}, \bibinfo {author} {\bibfnamefont {J.}~\bibnamefont {Armas}},
  \bibinfo {author} {\bibfnamefont {S.}~\bibnamefont {Bo}}, \bibinfo {author}
  {\bibfnamefont {C.}~\bibnamefont {Duclut}}, \bibinfo {author} {\bibfnamefont
  {F.}~\bibnamefont {J\"ulicher}},\ and\ \bibinfo {author} {\bibfnamefont
  {P.}~\bibnamefont {Sur\'owka}},\ }\bibfield  {title} {\bibinfo {title}
  {Passive odd viscoelasticity},\ }\href
  {https://doi.org/10.1103/PhysRevE.105.054607} {\bibfield  {journal} {\bibinfo
   {journal} {Phys. Rev. E}\ }\textbf {\bibinfo {volume} {105}},\ \bibinfo
  {pages} {054607} (\bibinfo {year} {2022})}\BibitemShut {NoStop}%
\bibitem [{\citenamefont {Sur\'owka}\ \emph {et~al.}(2023)\citenamefont
  {Sur\'owka}, \citenamefont {Souslov}, \citenamefont {J\"ulicher},\ and\
  \citenamefont {Banerjee}}]{surowka23}%
  \BibitemOpen
  \bibfield  {author} {\bibinfo {author} {\bibfnamefont {P.}~\bibnamefont
  {Sur\'owka}}, \bibinfo {author} {\bibfnamefont {A.}~\bibnamefont {Souslov}},
  \bibinfo {author} {\bibfnamefont {F.}~\bibnamefont {J\"ulicher}},\ and\
  \bibinfo {author} {\bibfnamefont {D.}~\bibnamefont {Banerjee}},\ }\bibfield
  {title} {\bibinfo {title} {Odd {Cosserat} elasticity in active materials},\
  }\href {https://doi.org/10.1103/PhysRevE.108.064609} {\bibfield  {journal}
  {\bibinfo  {journal} {Phys. Rev. E}\ }\textbf {\bibinfo {volume} {108}},\
  \bibinfo {pages} {064609} (\bibinfo {year} {2023})}\BibitemShut {NoStop}%
\bibitem [{\citenamefont {Shankar}\ and\ \citenamefont
  {Mahadevan}(2024)}]{shankar24}%
  \BibitemOpen
  \bibfield  {author} {\bibinfo {author} {\bibfnamefont {S.}~\bibnamefont
  {Shankar}}\ and\ \bibinfo {author} {\bibfnamefont {L.}~\bibnamefont
  {Mahadevan}},\ }\bibfield  {title} {\bibinfo {title} {Active hydraulics and
  odd elasticity of muscle fibres},\ }\href
  {https://doi.org/0.1038/s41567-024-02540-x} {\bibfield  {journal} {\bibinfo
  {journal} {Nat. Phys.}\ }\textbf {\bibinfo {volume} {20}},\ \bibinfo {pages}
  {1501} (\bibinfo {year} {2024})}\BibitemShut {NoStop}%
\bibitem [{\citenamefont {Zhang}\ and\ \citenamefont {Yao}(2024)}]{zhang24}%
  \BibitemOpen
  \bibfield  {author} {\bibinfo {author} {\bibfnamefont {Y.-H.}\ \bibnamefont
  {Zhang}}\ and\ \bibinfo {author} {\bibfnamefont {Z.}~\bibnamefont {Yao}},\
  }\bibfield  {title} {\bibinfo {title} {Anisotropic odd elasticity with
  {Hamiltonian} curl forces},\ }\href
  {https://doi.org/10.1088/1751-8121/ad8790} {\bibfield  {journal} {\bibinfo
  {journal} {J. Phys. A: Math. Theor.}\ }\textbf {\bibinfo {volume} {57}},\
  \bibinfo {pages} {455204} (\bibinfo {year} {2024})}\BibitemShut {NoStop}%
\bibitem [{\citenamefont {Lin}\ \emph {et~al.}(2024)\citenamefont {Lin},
  \citenamefont {Yasuda}, \citenamefont {Ishimoto},\ and\ \citenamefont
  {Komura}}]{lin24}%
  \BibitemOpen
  \bibfield  {author} {\bibinfo {author} {\bibfnamefont {L.-S.}\ \bibnamefont
  {Lin}}, \bibinfo {author} {\bibfnamefont {K.}~\bibnamefont {Yasuda}},
  \bibinfo {author} {\bibfnamefont {K.}~\bibnamefont {Ishimoto}},\ and\
  \bibinfo {author} {\bibfnamefont {S.}~\bibnamefont {Komura}},\ }\bibfield
  {title} {\bibinfo {title} {Emergence of odd elasticity in a microswimmer
  using deep reinforcement learning},\ }\href
  {https://doi.org/10.1103/PhysRevResearch.6.033016} {\bibfield  {journal}
  {\bibinfo  {journal} {Phys. Rev. Res.}\ }\textbf {\bibinfo {volume} {6}},\
  \bibinfo {pages} {033016} (\bibinfo {year} {2024})}\BibitemShut {NoStop}%
\bibitem [{\citenamefont {Walden}\ \emph {et~al.}(2025)\citenamefont {Walden},
  \citenamefont {Stegmaier}, \citenamefont {Dunkel},\ and\ \citenamefont
  {Mietke}}]{walden25}%
  \BibitemOpen
  \bibfield  {author} {\bibinfo {author} {\bibfnamefont {H.}~\bibnamefont
  {Walden}}, \bibinfo {author} {\bibfnamefont {A.}~\bibnamefont {Stegmaier}},
  \bibinfo {author} {\bibfnamefont {J.}~\bibnamefont {Dunkel}},\ and\ \bibinfo
  {author} {\bibfnamefont {A.}~\bibnamefont {Mietke}},\ }\bibfield  {title}
  {\bibinfo {title} {Odd electrical circuits},\ }\Eprint
  {https://arxiv.org/abs/2503.14383} {arXiv:2503.14383}  (\bibinfo {year}
  {2025})\BibitemShut {NoStop}%
\bibitem [{\citenamefont {Lee}\ \emph {et~al.}(2025)\citenamefont {Lee},
  \citenamefont {Lubensky},\ and\ \citenamefont {Markovich}}]{lee25}%
  \BibitemOpen
  \bibfield  {author} {\bibinfo {author} {\bibfnamefont {C.-T.}\ \bibnamefont
  {Lee}}, \bibinfo {author} {\bibfnamefont {T.~C.}\ \bibnamefont {Lubensky}},\
  and\ \bibinfo {author} {\bibfnamefont {T.}~\bibnamefont {Markovich}},\
  }\bibfield  {title} {\bibinfo {title} {Odd elasticity in disordered chiral
  active materials},\ }\Eprint {https://arxiv.org/abs/2508.04468}
  {arXiv:2508.04468}  (\bibinfo {year} {2025})\BibitemShut {NoStop}%
\bibitem [{\citenamefont {Zhou}\ \emph {et~al.}(2025)\citenamefont {Zhou},
  \citenamefont {Tsaloukidis}, \citenamefont {Binysh}, \citenamefont {Chen},
  \citenamefont {Fakhri}, \citenamefont {Coulais},\ and\ \citenamefont
  {Sur\'owka}}]{zhou25}%
  \BibitemOpen
  \bibfield  {author} {\bibinfo {author} {\bibfnamefont {Y.}~\bibnamefont
  {Zhou}}, \bibinfo {author} {\bibfnamefont {L.}~\bibnamefont {Tsaloukidis}},
  \bibinfo {author} {\bibfnamefont {J.}~\bibnamefont {Binysh}}, \bibinfo
  {author} {\bibfnamefont {Y.}~\bibnamefont {Chen}}, \bibinfo {author}
  {\bibfnamefont {N.}~\bibnamefont {Fakhri}}, \bibinfo {author} {\bibfnamefont
  {C.}~\bibnamefont {Coulais}},\ and\ \bibinfo {author} {\bibfnamefont
  {P.}~\bibnamefont {Sur\'owka}},\ }\bibfield  {title} {\bibinfo {title}
  {Curved odd elasticity},\ }\Eprint {https://arxiv.org/abs/2512.11037}
  {arXiv:2512.11037}  (\bibinfo {year} {2025})\BibitemShut {NoStop}%
\bibitem [{\citenamefont {Chao}\ \emph {et~al.}(2026)\citenamefont {Chao},
  \citenamefont {Gokhale}, \citenamefont {Lin}, \citenamefont {Hastewell},
  \citenamefont {Bacanu}, \citenamefont {Chen}, \citenamefont {Li},
  \citenamefont {Liu}, \citenamefont {Lee}, \citenamefont {Dunkel},\ and\
  \citenamefont {Fakri}}]{chao26}%
  \BibitemOpen
  \bibfield  {author} {\bibinfo {author} {\bibfnamefont {Y.-C.}\ \bibnamefont
  {Chao}}, \bibinfo {author} {\bibfnamefont {S.}~\bibnamefont {Gokhale}},
  \bibinfo {author} {\bibfnamefont {L.}~\bibnamefont {Lin}}, \bibinfo {author}
  {\bibfnamefont {A.}~\bibnamefont {Hastewell}}, \bibinfo {author}
  {\bibfnamefont {A.}~\bibnamefont {Bacanu}}, \bibinfo {author} {\bibfnamefont
  {Y.}~\bibnamefont {Chen}}, \bibinfo {author} {\bibfnamefont {J.}~\bibnamefont
  {Li}}, \bibinfo {author} {\bibfnamefont {J.}~\bibnamefont {Liu}}, \bibinfo
  {author} {\bibfnamefont {H.}~\bibnamefont {Lee}}, \bibinfo {author}
  {\bibfnamefont {J.}~\bibnamefont {Dunkel}},\ and\ \bibinfo {author}
  {\bibfnamefont {N.}~\bibnamefont {Fakri}},\ }\bibfield  {title} {\bibinfo
  {title} {Selective excitation of work-generating cycles in non-reciprocal
  living solids},\ }\href {https://doi.org/10.1038/s41567-026-03178-7}
  {\bibfield  {journal} {\bibinfo  {journal} {Nat. Phys.}\ }\textbf {\bibinfo
  {volume} {22}},\ \bibinfo {pages} {474} (\bibinfo {year} {2026})}\BibitemShut
  {NoStop}%
\bibitem [{\citenamefont {Bo}\ \emph {et~al.}(2026)\citenamefont {Bo},
  \citenamefont {Liu}, \citenamefont {He}, \citenamefont {Joyce}, \citenamefont
  {Leech}, \citenamefont {Tran}, \citenamefont {Ramirez}, \citenamefont
  {Boechler}, \citenamefont {Gravish}, \citenamefont {Zhao},\ and\
  \citenamefont {Tan}}]{bo26}%
  \BibitemOpen
  \bibfield  {author} {\bibinfo {author} {\bibfnamefont {F.}~\bibnamefont
  {Bo}}, \bibinfo {author} {\bibfnamefont {S.}~\bibnamefont {Liu}}, \bibinfo
  {author} {\bibfnamefont {Z.}~\bibnamefont {He}}, \bibinfo {author}
  {\bibfnamefont {W.}~\bibnamefont {Joyce}}, \bibinfo {author} {\bibfnamefont
  {G.}~\bibnamefont {Leech}}, \bibinfo {author} {\bibfnamefont
  {K.}~\bibnamefont {Tran}}, \bibinfo {author} {\bibfnamefont {K.}~\bibnamefont
  {Ramirez}}, \bibinfo {author} {\bibfnamefont {N.}~\bibnamefont {Boechler}},
  \bibinfo {author} {\bibfnamefont {N.}~\bibnamefont {Gravish}}, \bibinfo
  {author} {\bibfnamefont {H.}~\bibnamefont {Zhao}},\ and\ \bibinfo {author}
  {\bibfnamefont {T.~H.}\ \bibnamefont {Tan}},\ }\bibfield  {title} {\bibinfo
  {title} {Three phases of odd robotic active matter},\ }\Eprint
  {https://arxiv.org/abs/2603.09897} {arXiv:2603.09897}  (\bibinfo {year}
  {2026})\BibitemShut {NoStop}%
\bibitem [{\citenamefont {Mou}\ \emph {et~al.}(2026)\citenamefont {Mou},
  \citenamefont {Ren}, \citenamefont {Xu}, \citenamefont {Aranson},\ and\
  \citenamefont {Zhang}}]{Mou26}%
  \BibitemOpen
  \bibfield  {author} {\bibinfo {author} {\bibfnamefont {Z.}~\bibnamefont
  {Mou}}, \bibinfo {author} {\bibfnamefont {H.}~\bibnamefont {Ren}}, \bibinfo
  {author} {\bibfnamefont {D.}~\bibnamefont {Xu}}, \bibinfo {author}
  {\bibfnamefont {I.~S.}\ \bibnamefont {Aranson}},\ and\ \bibinfo {author}
  {\bibfnamefont {R.}~\bibnamefont {Zhang}},\ }\bibfield  {title} {\bibinfo
  {title} {Hydrodynamic modeling of odd nematic elasticity in liquid
  crystals},\ }\Eprint {https://arxiv.org/abs/2603.16977} {arXiv:2603.16977}
  (\bibinfo {year} {2026})\BibitemShut {NoStop}%
\bibitem [{\citenamefont {Lee}\ and\ \citenamefont {Markovich}(2026)}]{lee26}%
  \BibitemOpen
  \bibfield  {author} {\bibinfo {author} {\bibfnamefont {C.-T.}\ \bibnamefont
  {Lee}}\ and\ \bibinfo {author} {\bibfnamefont {T.}~\bibnamefont
  {Markovich}},\ }\bibfield  {title} {\bibinfo {title} {Non-hermitian chiral
  surface waves in disordered odd solids},\ }\Eprint
  {https://arxiv.org/abs/2603.21312} {arXiv:2603.21312}  (\bibinfo {year}
  {2026})\BibitemShut {NoStop}%
\bibitem [{\citenamefont {Binysh}\ \emph {et~al.}(2026)\citenamefont {Binysh},
  \citenamefont {Baardink}, \citenamefont {Veenstra}, \citenamefont {Coulais},\
  and\ \citenamefont {Souslov}}]{Binysh26}%
  \BibitemOpen
  \bibfield  {author} {\bibinfo {author} {\bibfnamefont {J.}~\bibnamefont
  {Binysh}}, \bibinfo {author} {\bibfnamefont {G.}~\bibnamefont {Baardink}},
  \bibinfo {author} {\bibfnamefont {J.}~\bibnamefont {Veenstra}}, \bibinfo
  {author} {\bibfnamefont {C.}~\bibnamefont {Coulais}},\ and\ \bibinfo {author}
  {\bibfnamefont {A.}~\bibnamefont {Souslov}},\ }\bibfield  {title} {\bibinfo
  {title} {More is less in unpercolated active solids},\ }\href
  {https://doi.org/10.1103/flhb-kjyd} {\bibfield  {journal} {\bibinfo
  {journal} {Phys. Rev. X}\ }\textbf {\bibinfo {volume} {16}},\ \bibinfo
  {pages} {021012} (\bibinfo {year} {2026})}\BibitemShut {NoStop}%
\bibitem [{\citenamefont {Shakespeare}(1599)}]{shakespeare}%
  \BibitemOpen
  \bibfield  {author} {\bibinfo {author} {\bibfnamefont {W.}~\bibnamefont
  {Shakespeare}},\ }\href@noop {} {\emph {\bibinfo {title} {The most exellent
  and lamentable tragedie, of Romeo and Iuliet}}}\ (\bibinfo  {publisher}
  {Cuthbert Burby},\ \bibinfo {address} {London, England},\ \bibinfo {year}
  {1599})\ \bibinfo {note} {{Act~2, Scene~2}}\BibitemShut {NoStop}%
\bibitem [{\citenamefont {Shakespeare}(2015)}]{rj}%
  \BibitemOpen
  \bibfield  {author} {\bibinfo {author} {\bibfnamefont {W.}~\bibnamefont
  {Shakespeare}},\ }\href@noop {} {\emph {\bibinfo {title} {Romeo and
  Juliet}}},\ edited by\ \bibinfo {editor} {\bibfnamefont {A.}~\bibnamefont
  {Poole}}\ (\bibinfo  {publisher} {Penguin},\ \bibinfo {address} {London,
  England},\ \bibinfo {year} {2015})\BibitemShut {NoStop}%
\bibitem [{\citenamefont {Avron}(1998)}]{avron98}%
  \BibitemOpen
  \bibfield  {author} {\bibinfo {author} {\bibfnamefont {J.~E.}\ \bibnamefont
  {Avron}},\ }\bibfield  {title} {\bibinfo {title} {Odd viscosity},\ }\href
  {https://doi.org/10.1023/A:1023084404080} {\bibfield  {journal} {\bibinfo
  {journal} {J. Stat. Phys.}\ }\textbf {\bibinfo {volume} {92}},\ \bibinfo
  {pages} {543} (\bibinfo {year} {1998})}\BibitemShut {NoStop}%
\bibitem [{\citenamefont {Rivlin}(1974)}]{rivlin_cube}%
  \BibitemOpen
  \bibfield  {author} {\bibinfo {author} {\bibfnamefont {R.~S.}\ \bibnamefont
  {Rivlin}},\ }\bibfield  {title} {\bibinfo {title} {Stability of pure
  homogeneous deformations of an elastic cube under dead loading},\ }\href
  {https://doi.org/doi.org/10.1090/qam/99680} {\bibfield  {journal} {\bibinfo
  {journal} {Quart. Appl. Math.}\ }\textbf {\bibinfo {volume} {32}},\ \bibinfo
  {pages} {265} (\bibinfo {year} {1974})}\BibitemShut {NoStop}%
\bibitem [{\citenamefont {Ball}\ and\ \citenamefont
  {Schaeffer}(1983)}]{ball1983}%
  \BibitemOpen
  \bibfield  {author} {\bibinfo {author} {\bibfnamefont {J.~M.}\ \bibnamefont
  {Ball}}\ and\ \bibinfo {author} {\bibfnamefont {D.~G.}\ \bibnamefont
  {Schaeffer}},\ }\bibfield  {title} {\bibinfo {title} {Bifurcation and
  stability of homogeneous equilibrium configurations of an elastic body under
  dead-load tractions},\ }\href {https://doi.org/10.1017/S030500410006117X}
  {\bibfield  {journal} {\bibinfo  {journal} {Math. Proc. Camb. Philos. Soc.}\
  }\textbf {\bibinfo {volume} {94}},\ \bibinfo {pages} {315} (\bibinfo {year}
  {1983})}\BibitemShut {NoStop}%
\bibitem [{\citenamefont {Goriely}(2017)}]{goriely}%
  \BibitemOpen
  \bibfield  {author} {\bibinfo {author} {\bibfnamefont {A.}~\bibnamefont
  {Goriely}},\ }\href@noop {} {\emph {\bibinfo {title} {The Mathematics and
  Mechanics of Biological Growth}}}\ (\bibinfo  {publisher} {Springer},\
  \bibinfo {address} {Berlin, Germany},\ \bibinfo {year} {2017})\
  Chap.~\bibinfo {chapter} {11}, pp.\ \bibinfo {pages} {261--344}\BibitemShut
  {NoStop}%
\bibitem [{Note1()}]{Note1}%
  \BibitemOpen
  \bibinfo {note} {In hyperelasticity~\cite {goriely}, an analogous argument
  establishes that the energy density $W$ is a (scalar) isotropic function of
  $\protect \mathbfsfit {B}$, but we are not aware of a reference for this
  result in Cauchy elasticity.}\BibitemShut {Stop}%
\bibitem [{\citenamefont {Zheng}(1994)}]{zheng94}%
  \BibitemOpen
  \bibfield  {author} {\bibinfo {author} {\bibfnamefont {Q.-S.}\ \bibnamefont
  {Zheng}},\ }\bibfield  {title} {\bibinfo {title} {Theory of representations
  for tensor functions - a unified invariant approach to constitutive
  equations},\ }\href {https://doi.org/10.1115/1.3111066} {\bibfield  {journal}
  {\bibinfo  {journal} {Appl. Mech. Rev.}\ }\textbf {\bibinfo {volume} {47}},\
  \bibinfo {pages} {545} (\bibinfo {year} {1994})}\BibitemShut {NoStop}%
\bibitem [{Note2()}]{Note2}%
  \BibitemOpen
  \bibinfo {note} {The corresponding result in 3D hyperelasticity is
  well-known~\cite {goriely,rivlin97}, but we are not aware of an exact
  reference for the 2D result: since the energy density $W$ is a scalar
  function of $\protect \mathbfsfit {B}$ only~\cite {Note1}, it is a function
  of $\protect \mathcal {I},\protect \mathcal {J}$ only; the definition
  $\protect \mathbfsfit {T}=\protect \mathcal {J}^{-1}\protect \mathbfsfit
  {F}\partial W/\partial \protect \mathbfsfit {F}$~\cite {goriely} then yields
  the representation $\protect \mathbfsfit {T}=t_1\protect \mathbfsfit
  {I}+t_2\protect \mathbfsfit {B}$, where $t_1=\partial W/\partial \protect
  \mathcal {J}$ and $t_2=2\protect \mathcal {J}^{-1}\partial W/\partial
  \protect \mathcal {I}$. Thus, unlike in Eq.~\protect \hyperref [eq:T1]{(\ref
  *{eq:T1})}, $t_1,t_2$ are not arbitrary. This means that $2\partial
  t_1/\partial \protect \mathcal {I} \protect \neq t_2+\protect \mathcal
  {J}\partial t_2/\partial \protect \mathcal {J}$ introduces odd behaviour,
  too. We will raise a similar point in our discussion of 3D nonlinear odd
  elasticity.}\BibitemShut {Stop}%
\bibitem [{\citenamefont {Rivlin}\ and\ \citenamefont
  {Ericksen}(1997)}]{rivlin97}%
  \BibitemOpen
  \bibfield  {author} {\bibinfo {author} {\bibfnamefont {R.~S.}\ \bibnamefont
  {Rivlin}}\ and\ \bibinfo {author} {\bibfnamefont {J.~L.}\ \bibnamefont
  {Ericksen}},\ }\bibinfo {title} {Stress-deformation relations for isotropic
  materials},\ in\ \href {https://doi.org/10.1007/978-1-4612-2416-7_61} {\emph
  {\bibinfo {booktitle} {Collected Papers of R.S. Rivlin: Volume I and II}}},\
  \bibinfo {editor} {edited by\ \bibinfo {editor} {\bibfnamefont {G.~I.}\
  \bibnamefont {Barenblatt}}\ and\ \bibinfo {editor} {\bibfnamefont {D.~D.}\
  \bibnamefont {Joseph}}}\ (\bibinfo  {publisher} {Springer New York},\
  \bibinfo {address} {New York, NY},\ \bibinfo {year} {1997})\ pp.\ \bibinfo
  {pages} {911--1013}\BibitemShut {NoStop}%
\bibitem [{Note3()}]{Note3}%
  \BibitemOpen
  \bibinfo {note} {See Supplemental Material at [url to be inserted], which
  includes Refs.~\cite
  {surowka24,Chen2017,Zhao2021,ogden,goriely,Overvelde2016,haas21,yavari2025},
  for details of the calculations.}\BibitemShut {Stop}%
\bibitem [{\citenamefont {Ostoja-Starzewski}\ and\ \citenamefont
  {Sur\'owka}(2024)}]{surowka24}%
  \BibitemOpen
  \bibfield  {author} {\bibinfo {author} {\bibfnamefont {M.}~\bibnamefont
  {Ostoja-Starzewski}}\ and\ \bibinfo {author} {\bibfnamefont {P.}~\bibnamefont
  {Sur\'owka}},\ }\bibfield  {title} {\bibinfo {title} {Generalizing odd
  elasticity theory to odd thermoelasticity for planar materials},\ }\href
  {https://doi.org/10.1103/PhysRevB.109.064107} {\bibfield  {journal} {\bibinfo
   {journal} {Phys. Rev. B}\ }\textbf {\bibinfo {volume} {109}},\ \bibinfo
  {pages} {064107} (\bibinfo {year} {2024})}\BibitemShut {NoStop}%
\bibitem [{\citenamefont {Chen}\ \emph {et~al.}(2017)\citenamefont {Chen},
  \citenamefont {Chang},\ and\ \citenamefont {Qin}}]{Chen2017}%
  \BibitemOpen
  \bibfield  {author} {\bibinfo {author} {\bibfnamefont {L.}~\bibnamefont
  {Chen}}, \bibinfo {author} {\bibfnamefont {Z.}~\bibnamefont {Chang}},\ and\
  \bibinfo {author} {\bibfnamefont {T.}~\bibnamefont {Qin}},\ }\bibfield
  {title} {\bibinfo {title} {Elastic wave propagation in simple-sheared
  hyperelastic materials with different constitutive models},\ }\href
  {https://doi.org/https://doi.org/10.1016/j.ijsolstr.2017.07.027} {\bibfield
  {journal} {\bibinfo  {journal} {Int. J. Solids Struct.}\ }\textbf {\bibinfo
  {volume} {126--127}},\ \bibinfo {pages} {1} (\bibinfo {year}
  {2017})}\BibitemShut {NoStop}%
\bibitem [{\citenamefont {Zhao}\ and\ \citenamefont {Chang}(2021)}]{Zhao2021}%
  \BibitemOpen
  \bibfield  {author} {\bibinfo {author} {\bibfnamefont {S.}~\bibnamefont
  {Zhao}}\ and\ \bibinfo {author} {\bibfnamefont {Z.}~\bibnamefont {Chang}},\
  }\bibfield  {title} {\bibinfo {title} {Elastic wave velocities in finitely
  pre-stretched soft fibers},\ }\href
  {https://doi.org/https://doi.org/10.1016/j.ijsolstr.2021.111208} {\bibfield
  {journal} {\bibinfo  {journal} {Int. J. Solids Struct.}\ }\textbf {\bibinfo
  {volume} {233}},\ \bibinfo {pages} {111208} (\bibinfo {year}
  {2021})}\BibitemShut {NoStop}%
\bibitem [{\citenamefont {Overvelde}\ \emph {et~al.}(2016)\citenamefont
  {Overvelde}, \citenamefont {Dykstra}, \citenamefont {de~Rooij}, \citenamefont
  {Weaver},\ and\ \citenamefont {Bertoldi}}]{Overvelde2016}%
  \BibitemOpen
  \bibfield  {author} {\bibinfo {author} {\bibfnamefont {J.~T.~B.}\
  \bibnamefont {Overvelde}}, \bibinfo {author} {\bibfnamefont {D.~M.~J.}\
  \bibnamefont {Dykstra}}, \bibinfo {author} {\bibfnamefont {R.}~\bibnamefont
  {de~Rooij}}, \bibinfo {author} {\bibfnamefont {J.}~\bibnamefont {Weaver}},\
  and\ \bibinfo {author} {\bibfnamefont {K.}~\bibnamefont {Bertoldi}},\
  }\bibfield  {title} {\bibinfo {title} {Tensile instability in a thick elastic
  body},\ }\href {https://doi.org/10.1103/PhysRevLett.117.094301} {\bibfield
  {journal} {\bibinfo  {journal} {Phys. Rev. Lett.}\ }\textbf {\bibinfo
  {volume} {117}},\ \bibinfo {pages} {094301} (\bibinfo {year}
  {2016})}\BibitemShut {NoStop}%
\bibitem [{\citenamefont {Haas}\ and\ \citenamefont
  {Goldstein}(2021)}]{haas21}%
  \BibitemOpen
  \bibfield  {author} {\bibinfo {author} {\bibfnamefont {P.~A.}\ \bibnamefont
  {Haas}}\ and\ \bibinfo {author} {\bibfnamefont {R.~E.}\ \bibnamefont
  {Goldstein}},\ }\bibfield  {title} {\bibinfo {title} {Morphoelasticity of
  large bending deformations of cell sheets during development},\ }\href
  {https://doi.org/10.1103/PhysRevE.103.022411} {\bibfield  {journal} {\bibinfo
   {journal} {Phys. Rev. E}\ }\textbf {\bibinfo {volume} {103}},\ \bibinfo
  {pages} {022411} (\bibinfo {year} {2021})}\BibitemShut {NoStop}%
\bibitem [{vno()}]{vnote}%
  \BibitemOpen
  \href@noop {} {}\bibinfo {note} {We have chosen the simplest volumetric
  penalty, $\mathcal{J}-1$, here; other choices (e.g., $\log{\mathcal{J}}$) are
  common, too, but our calculations~\cite{Note3} show that this does not affect
  the subsequent results.}\BibitemShut {Stop}%
\bibitem [{\citenamefont {Rivlin}\ and\ \citenamefont
  {Beatty}(2003)}]{rivlin2003}%
  \BibitemOpen
  \bibfield  {author} {\bibinfo {author} {\bibfnamefont {R.~S.}\ \bibnamefont
  {Rivlin}}\ and\ \bibinfo {author} {\bibfnamefont {M.~F.}\ \bibnamefont
  {Beatty}},\ }\bibfield  {title} {\bibinfo {title} {Dead loading of a unit
  cube of compressible isotropic elastic material},\ }\href
  {https://doi.org/10.1007/s00033-003-3070-z} {\bibfield  {journal} {\bibinfo
  {journal} {Z. angew. Math. Phys.}\ }\textbf {\bibinfo {volume} {54}},\
  \bibinfo {pages} {954} (\bibinfo {year} {2003})}\BibitemShut {NoStop}%
\bibitem [{\citenamefont {Haughton}(2005)}]{haughton2005}%
  \BibitemOpen
  \bibfield  {author} {\bibinfo {author} {\bibfnamefont {D.~M.}\ \bibnamefont
  {Haughton}},\ }\bibfield  {title} {\bibinfo {title} {A comparison of
  stability and bifurcation criteria for a compressible elastic cube},\ }\href
  {https://doi.org/10.1007/s10665-005-4752-7} {\bibfield  {journal} {\bibinfo
  {journal} {J. Eng. Math.}\ }\textbf {\bibinfo {volume} {53}},\ \bibinfo
  {pages} {79} (\bibinfo {year} {2005})}\BibitemShut {NoStop}%
\bibitem [{\citenamefont {Tarantino}(2008)}]{tarantino2008}%
  \BibitemOpen
  \bibfield  {author} {\bibinfo {author} {\bibfnamefont {A.~M.}\ \bibnamefont
  {Tarantino}},\ }\bibfield  {title} {\bibinfo {title} {Homogeneous equilibrium
  configurations of a hyperelastic compressible cube under equitriaxial
  dead-load tractions},\ }\href {https://doi.org/10.1007/s10659-008-9160-6}
  {\bibfield  {journal} {\bibinfo  {journal} {J. Elast.}\ }\textbf {\bibinfo
  {volume} {92}},\ \bibinfo {pages} {227} (\bibinfo {year} {2008})}\BibitemShut
  {NoStop}%
\bibitem [{\citenamefont {Mihai}\ \emph {et~al.}(2019)\citenamefont {Mihai},
  \citenamefont {Woolley},\ and\ \citenamefont {Goriely}}]{Mihai2019}%
  \BibitemOpen
  \bibfield  {author} {\bibinfo {author} {\bibfnamefont {L.~A.}\ \bibnamefont
  {Mihai}}, \bibinfo {author} {\bibfnamefont {T.~E.}\ \bibnamefont {Woolley}},\
  and\ \bibinfo {author} {\bibfnamefont {A.}~\bibnamefont {Goriely}},\
  }\bibfield  {title} {\bibinfo {title} {Likely equilibria of the stochastic
  {Rivlin} cube},\ }\href {https://doi.org/10.1098/rsta.2018.0068} {\bibfield
  {journal} {\bibinfo  {journal} {Phil. Trans. R. Soc. A}\ }\textbf {\bibinfo
  {volume} {377}},\ \bibinfo {pages} {20180068} (\bibinfo {year}
  {2019})}\BibitemShut {NoStop}%
\bibitem [{Note4()}]{Note4}%
  \BibitemOpen
  \bibinfo {note} {The Rivlin problem is a solvable paradigm of nonlinear
  elasticity~\cite {Mihai2019}, but experimental realisations are tricky, if
  not impossible~\cite {Overvelde2016} because, with the assumed dead loads,
  the total force on each side of $\protect \mathcal {S}$ is constant, so the
  (Cauchy) stress on the deformed faces is not constant.}\BibitemShut {Stop}%
\bibitem [{\citenamefont {Roychowdhury}\ \emph {et~al.}(2025)\citenamefont
  {Roychowdhury}, \citenamefont {Rao},\ and\ \citenamefont
  {Truskinovsky}}]{Roychowdhury2025}%
  \BibitemOpen
  \bibfield  {author} {\bibinfo {author} {\bibfnamefont {A.}~\bibnamefont
  {Roychowdhury}}, \bibinfo {author} {\bibfnamefont {M.}~\bibnamefont {Rao}},\
  and\ \bibinfo {author} {\bibfnamefont {L.}~\bibnamefont {Truskinovsky}},\
  }\bibfield  {title} {\bibinfo {title} {Active drive towards elastic
  spinodals},\ }\href {https://doi.org/10.1103/h2yp-d2s2} {\bibfield  {journal}
  {\bibinfo  {journal} {Phys. Rev. E}\ }\textbf {\bibinfo {volume} {111}},\
  \bibinfo {pages} {065416} (\bibinfo {year} {2025})}\BibitemShut {NoStop}%
\bibitem [{\citenamefont {Landau}\ and\ \citenamefont
  {Lifshitz}(1986)}]{landaulifshitz}%
  \BibitemOpen
  \bibfield  {author} {\bibinfo {author} {\bibfnamefont {L.~D.}\ \bibnamefont
  {Landau}}\ and\ \bibinfo {author} {\bibfnamefont {E.~M.}\ \bibnamefont
  {Lifshitz}},\ }\href@noop {} {\emph {\bibinfo {title} {Theory of
  Elasticity}}},\ \bibinfo {edition} {3rd}\ ed.,\ \bibinfo {series} {Course of
  Theoretical Physics}, Vol.~\bibinfo {volume} {7}\ (\bibinfo  {publisher}
  {Elsevier},\ \bibinfo {address} {Oxford, United Kingdom},\ \bibinfo {year}
  {1986})\ Chap.~\bibinfo {chapter} {4}, pp.\ \bibinfo {pages}
  {9--11}\BibitemShut {NoStop}%
\bibitem [{\citenamefont {Boas}(2006)}]{boas2006}%
  \BibitemOpen
  \bibfield  {author} {\bibinfo {author} {\bibfnamefont {M.~L.}\ \bibnamefont
  {Boas}},\ }\href@noop {} {\emph {\bibinfo {title} {Mathematical Methods in
  the Physical Sciences}}},\ \bibinfo {edition} {3rd}\ ed.\ (\bibinfo
  {publisher} {John Wiley \& Sons},\ \bibinfo {address} {Hoboken, NJ},\
  \bibinfo {year} {2006})\ Chap.\ \bibinfo {chapter} {6.11}, pp.\ \bibinfo
  {pages} {324--336}\BibitemShut {NoStop}%
\bibitem [{\citenamefont {Rivlin}\ \emph {et~al.}(1951)\citenamefont {Rivlin},
  \citenamefont {Saunders},\ and\ \citenamefont {Andrade}}]{rivlin1951}%
  \BibitemOpen
  \bibfield  {author} {\bibinfo {author} {\bibfnamefont {R.~S.}\ \bibnamefont
  {Rivlin}}, \bibinfo {author} {\bibfnamefont {D.~W.}\ \bibnamefont
  {Saunders}},\ and\ \bibinfo {author} {\bibfnamefont {E.~N. D.~C.}\
  \bibnamefont {Andrade}},\ }\bibfield  {title} {\bibinfo {title} {Large
  elastic deformations of isotropic materials {VII. Experiments} on the
  deformation of rubber},\ }\href {https://doi.org/10.1098/rsta.1951.0004}
  {\bibfield  {journal} {\bibinfo  {journal} {Phil. Trans. R. Soc. A}\ }\textbf
  {\bibinfo {volume} {243}},\ \bibinfo {pages} {251} (\bibinfo {year}
  {1951})}\BibitemShut {NoStop}%
\bibitem [{\citenamefont {Zhao}\ and\ \citenamefont {Haas}(2025)}]{zhao2025}%
  \BibitemOpen
  \bibfield  {author} {\bibinfo {author} {\bibfnamefont {S.}~\bibnamefont
  {Zhao}}\ and\ \bibinfo {author} {\bibfnamefont {P.~A.}\ \bibnamefont
  {Haas}},\ }\bibfield  {title} {\bibinfo {title} {Mechanics of poking a
  cyst},\ }\href {https://doi.org/10.1103/PhysRevLett.134.228402} {\bibfield
  {journal} {\bibinfo  {journal} {Phys. Rev. Lett.}\ }\textbf {\bibinfo
  {volume} {134}},\ \bibinfo {pages} {228402} (\bibinfo {year}
  {2025})}\BibitemShut {NoStop}%
\end{thebibliography}%
\end{document}


\renewcommand{\theequation}{S\arabic{equation}}
\renewcommand{\thefigure}{S\arabic{figure}}
\renewcommand{\thepage}{S\arabic{page}}

\begin{CJK*}{UTF8}{gbsn}
\title{Nonlinear isotropic odd elasticity\\\textbullet{} Supplemental Material \textbullet{}}

\author{Shiheng Zhao (赵世恒)}
\affiliation{Max Planck Institute for the Physics of Complex Systems, N\"othnitzer Stra\ss e 38, 01187 Dresden, Germany}
\affiliation{\mbox{Max Planck Institute of Molecular Cell Biology and Genetics, Pfotenhauerstra\ss e 108, 01307 Dresden, Germany}}
\affiliation{Center for Systems Biology Dresden, Pfotenhauerstra\ss e 108, 01307 Dresden, Germany}
\author{Pierre A. Haas}
\affiliation{Max Planck Institute for the Physics of Complex Systems, N\"othnitzer Stra\ss e 38, 01187 Dresden, Germany}
\affiliation{\mbox{Max Planck Institute of Molecular Cell Biology and Genetics, Pfotenhauerstra\ss e 108, 01307 Dresden, Germany}}
\affiliation{Center for Systems Biology Dresden, Pfotenhauerstra\ss e 108, 01307 Dresden, Germany}

\maketitle
\end{CJK*}

This Supplemental Material provides the details of the calculations in the main text.

\section{\uppercase{Isotropic nonlinear odd elasticity in 2D}}
\subsection{Linearised constitutive relations}
We begin by linearising the constitutive relations of 2D isotropic nonlinear odd elasticity, 
\begin{align}
\tens{T}=t_1(\mathcal{I},\mathcal{J})\tens{I}+t_2(\mathcal{I},\mathcal{J})\dev{\tens{B}}+t_3(\mathcal{I},\mathcal{J})[\tens{B},\teps]\!+\!t_4(\mathcal{I},\mathcal{J})\teps,\label{eq:four_term_decomp_main}
\end{align}
as in Eq.~(1b) of the main text, about the undeformed reference state. As announced in the main text, we let $\tens{F}=\tens{I}+\vec{\nabla u}$, where $\|\vec{\nabla u}\|\ll 1$. We define $\tens{S}=\bigl(\vec{\nabla u}+\vec{\nabla u}^\top\bigr)/2$. Then, to first order,
\begin{subequations}\label{eq:lin_B_J_main}
\begin{align}
\tens{B}&=\tens{F}\tens{F}^\top=\tens{I}+2\tens{S}+\mathcal{O}\bigl(\|\vec{\nabla u}\|^2\bigr),\\
\mathcal{J}&=\det\tens{F}=1+\vabla\cdot\vec{ u} +\mathcal{O}\bigl(\|\vec{\nabla u}\|^2\bigr),\\
\mathcal{I}&=\tr{\tens{B}}= 2 + 2\vabla\cdot\vec{ u}+\mathcal{O}\bigl(\|\vec{\nabla u}\|^2\bigr).
\end{align}
\end{subequations}
Thence
\begin{subequations}\label{eq:lin_dev_comm_main}
\begin{align}
\dev{\tens{B}}&=2\dev{\tens{S}} +\mathcal{O}\bigl(\|\vec{\nabla u}\|^2\bigr),\\
[\tens{B},\teps]&=2[\tens{S},\teps]+\mathcal{O}\bigl(\|\vec{\nabla u}\|^2\bigr),
\end{align}
\end{subequations}
since $\dev{\tens{I}}=[\tens{I},\teps]=\tzero$. We now expand each scalar coefficient in Eq.~\eqref{eq:four_term_decomp_main} about $\mathcal I=2$, $\mathcal J=1$, viz.,
\begin{align}
t_i(\mathcal{I}, \mathcal{J}) &= t_i(2, 1) + (\mathcal{I} - 2)\left.\frac{\partial t_i}{\partial \mathcal{I}}\right|_{\substack{\mathcal{I}=2\\\mathcal{J}=1}}  + (\mathcal{J} - 1)\left. \frac{\partial t_i}{\partial \mathcal{J}} \right|_{\substack{\mathcal{I}=2\\\mathcal{J}=1}}\nonumber\\
&\quad+\mathcal{O}\bigl(\|\vec{\nabla u}\|^2\bigr),
\label{eq:taylor_ta_main}
\end{align}
for $i=1,2,3,4$. To enforce a stress-free reference state at $\tens{F}=\tens{I}$, we need to impose 
\begin{align}
&t_1(2,1)=0, &&t_4(2,1)=0,
\label{eq:reference_zero_main}
\end{align}
as announced in the main text. Substituting Eqs.~\eqref{eq:lin_B_J_main}--\eqref{eq:taylor_ta_main} into Eq.~\eqref{eq:four_term_decomp_main} gives $\tens{T}=\tens{t}+\mathcal{O}\bigl(\|\vec{\nabla u}\|^2\bigr)$, with
\begin{align}
\tens{t} &=\left(2\frac{\partial t_1}{\partial \mathcal{I}} + \frac{\partial t_1}{\partial \mathcal{J}}\right)(\vabla\cdot\vec{ u})\,\tens{I}+2t_2\,\dev{\tens{S}}+2t_3[\tens{S},\teps]\nonumber\\
&\qquad+\left(2\frac{\partial t_4}{\partial \mathcal{I}}+\frac{\partial t_4}{\partial \mathcal{J}}\right)(\vabla\cdot\vec{ u})\,\teps,
\label{eq:dleq_general_main}
\end{align}
in which $t_1,t_2,t_3,t_4$ and their derivatives are evaluated at $\mathcal I=2$, $\mathcal J=1$. This is Eq.~(2b) of the main text, with
\begin{align}
&t_1^{(1)}\!=2\frac{\partial t_1}{\partial \mathcal{I}}\!+\!\frac{\partial t_1}{\partial \mathcal{J}},&&t_2^{(0)}\!=\!2t_2,&&t_3^{(0)}\!=\!2t_3,&&t_4^{(1)}\!=2\frac{\partial t_4}{\partial \mathcal{I}}\!+\!\frac{\partial t_4}{\partial \mathcal{J}}.
\end{align}
We compare Eq.~\eqref{eq:dleq_general_main} with the constitutive relations of isotropic linear odd elasticity~\cite{surowka24}, ${\tens{t}=\tens{C}:\vec{\nabla u}}$, where
\begin{align}
C_{ijk\ell} &=\kappa\delta_{ij}\delta_{k\ell}+A\,\epsilon_{ij}\delta_{k\ell}+\mu\bigl(\delta_{ik}\delta_{j\ell}+\delta_{i\ell}\delta_{jk}-\delta_{ij}\delta_{k\ell}\bigr)\nonumber\\
&\qquad-K\bigl(\epsilon_{ik}\delta_{j\ell}+\epsilon_{j\ell}\delta_{ik}\bigr).
\label{eq:C_linear_odd_main}
\end{align}
Explicitly,
\begin{align}
\tens{t}=\kappa(\vabla\cdot\vec{ u})\tens{I}-A(\vabla\cdot\vec{ u})\teps+2\mu\dev{\tens{S}}-K[\tens{S},\teps].
\label{eq:sigma_linear_main}
\end{align}
Matching Eqs.~\eqref{eq:dleq_general_main} and \eqref{eq:sigma_linear_main} yields the constraints
\begin{subequations}
\begin{align}
&t_2^{(0)}=t_2(2,1)=2\mu,&& t_3^{(0)}=t_3(2,1)=-K,
\label{eq:match_shear_main}
\end{align}
and
\begin{align}
t_1^{(1)}&=2\frac{\partial t_1}{\partial \mathcal{I}}+\frac{\partial t_1}{\partial \mathcal{J}}=\kappa, &
t_4^{(1)}&=2\frac{\partial t_4}{\partial \mathcal{I}} + \frac{\partial t_4}{\partial \mathcal{J}}=-A.
\label{eq:match_bulk_main}
\end{align}
\end{subequations}
Thus, linear odd elasticity fixes the values of $t_2,t_3$ at the reference state and the directional derivatives of $t_1,t_4$ along the line $\mathcal{I}=2\mathcal{J}$. This stresses the fact that there is no unique nonlinear completion of the constitutive relations of linear odd elasticity, as expected.

\subsection{Minimal model of isotropic nonlinear odd elasticity in 2D}
As announced in the main text, we choose a minimal completion that satisfies the following requirements: (i) the reference state is stress-free, (ii) the coefficients have no explicit dependence on $\mathcal{I}$, (iii) the bulk and shear responses inherit their dependence on $\mathcal{J}$ from 2D neo-Hookean elasticity~\cite{Chen2017,Zhao2021}. This implies
\begin{subequations}\label{eq:minimal_completion_main}
\begin{align}
t_2(\mathcal I,\mathcal J)&=2\mu\,\mathcal J^{-2}, &t_3(\mathcal I,\mathcal J)&=-K\mathcal J^{-2}, \\
t_1(\mathcal I,\mathcal J)&=\kappa(\mathcal J-1),&t_4(\mathcal I,\mathcal J)&=-A(\mathcal J-1),
\end{align}
\end{subequations}
for the shear and bulk responses, respectively. This choice clearly satisfies Eqs.~\eqref{eq:reference_zero_main}, \eqref{eq:match_shear_main}, and \eqref{eq:match_bulk_main}. Substituting Eqs.~\eqref{eq:minimal_completion_main} into Eq.~\eqref{eq:four_term_decomp_main} yields
\begin{align}
\tens{T}=\kappa(\mathcal{J}-1)\tens{I}+\dfrac{\mu}{\mathcal{J}^2}\dev{\tens{B}}-\dfrac{K}{2\mathcal{J}^2}[\tens{B},\teps]-A(\mathcal{J}-1)\teps,\label{eq:comp_T}
\end{align}
which is Eq.~(4a) of the main text. As explained in the main text, the corresponding incompressible stress tensor is
\begin{align}
\tens{T}=-p\tens{I}+\mu\tens{B}-\frac{K}{2}[\tens{B},\teps],
\label{eq:incomp_T}
\end{align}
as in Eq.~(4b) of the main text.

\section{\uppercase{The compressible odd Rivlin square}}
As discussed in the main text, we study the deformations of a square under the homogeneous dead-load deformation $\tens{P}=\tau \tens{I}$, where $\tau$ is the nominal dead load. It will be useful to note that rotation by $\theta$ is
\begin{align}
\label{eq:Rtheta_square}
\mathcalbf{R}(\theta)=\cos\theta\,\tens{I}+\sin\theta\,\teps\in \text{SO}(2).
\end{align}

\subsection{The square branch}
Because the loading is symmetric and the constitutive law is isotropic, we expect a trivial, square branch in which the deformation is an isotropic stretch by $\lambda>0$ followed by a rotation,
\begin{align}
&\tens{F}=\lambda \tens{R}, &&\text{with }\tens{R}\in \text{SO}(2).\label{eq:Fsym}
\end{align}
This implies $\mathcal{J}=\lambda^2$, $\tens{B}=\lambda^2\tens{I}$, so $\dev{\tens{B}}=\tzero$, $[\tens{B},\teps]=\tzero$. Equation~\eqref{eq:comp_T} then gives $\tens{T}=(\lambda^2-1)(\kappa \tens{I}-A\teps)$. Since $\tens{F}^{-\top}=\lambda^{-1}\tens{R}$, the first Piola--Kirchhoff stress is
\begin{align}
\label{eq:Psq_pre}
\tens{P} =\mathcal{J}\tens{TF}^{-\top}=\lambda(\lambda^2-1)(\kappa \tens{I}-A\teps)\tens{R}.
\end{align}
Imposing $\tens{P}=\tau\tens{I}$ implies
\begin{align}
\tens{R}=\dfrac{\tau}{\lambda(\lambda^2-1)}(\kappa\tens{I}-A\teps)^{-1}.
\end{align}
Let $\vartheta_0$ be such that $\tan{\vartheta_0}=A/\kappa$. Then
\begin{align}
\mathcalbf{R}(-\vartheta_0)=\dfrac{1}{\sqrt{\kappa^2+A^2}}(\kappa\tens{I}-A\teps),
\end{align}
so $(\kappa\tens{I}-A\teps)^{-1}\!=\!(\kappa^2+A^2)^{-1/2}\mathcalbf{R}(-\vartheta_0)^{-1}\!=\!(\kappa^2+A^2)^{-1/2}\mathcalbf{R}(\vartheta_0)$. Thus
\begin{align}
\tens{R}=\dfrac{\tau}{\lambda(\lambda^2-1)\sqrt{\kappa^2+A^2}}\mathcalbf{R}(\vartheta_0).
\end{align}
Since $\tens{R}\in\text{SO}(2)$, this implies
\begin{align}
\tens{R}=\mathcalbf{R}(\vartheta_0)\quad\text{and}\quad\tau=\lambda(\lambda^2-1)\sqrt{\kappa^2+A^2}.\label{eq:symsol}
\end{align}
This determines the square branch. In particular, the odd volumetric term only selects a fixed rotation.

\subsection{Symmetry-breaking of the square branch}
To seek homogeneous deformation states with broken symmetry, we consider the more general uniform ansatz
\begin{align}
\label{eq:F_general_uniform_square}
&\tens{F}=\tens{V}\tens{R}, &&\text{with }\tens{R}\in\text{SO}(2),\;\tens{V}=\operatorname{diag}(\lambda_1,\lambda_2),
\end{align}
where $\lambda_1,\lambda_2>0$ are principal stretches. With this, $\mathcal{J}=\lambda_1\lambda_2$, $\tens{B}=\operatorname{diag}\bigl(\lambda_1^2, \lambda_2^2\bigr)$. Direct computation gives
\begin{subequations}
\begin{align}
\dev{\tens{B}} &=\frac{1}{2}
\begin{pmatrix}
\lambda_1^2-\lambda_2^2 & 0\\
0 & \lambda_2^2-\lambda_1^2
\end{pmatrix},\\ 
[\tens{B},\teps] &=\bigl(\lambda_2^2-\lambda_1^2\bigr)
\begin{pmatrix}
0 & 1\\
1 & 0
\end{pmatrix}.
\end{align}
\end{subequations}
Hence, from Eq.~\eqref{eq:comp_T}, the components of the Cauchy stress are
\begin{subequations}\label{eq:T_general_uniform_square}
\begin{align}
    T_{11} &= \kappa(\mathcal{J}-1)+\dfrac{\mu}{2\mathcal{J}^2}\bigl(\lambda_1^2-\lambda_2^2\bigr), \\
    T_{12} &= A(\mathcal{J}-1)+\dfrac{K}{2\mathcal{J}^2}\bigl(\lambda_1^2-\lambda_2^2\bigr),\\
    T_{21} &= -A(\mathcal{J}-1)+\dfrac{K}{2\mathcal{J}^{2}}\bigl(\lambda_1^2-\lambda_2^2\bigr),\\
    T_{22} &= \kappa(\mathcal{J}-1)-\dfrac{\mu}{2\mathcal{J}^2}\bigl(\lambda_1^2-\lambda_2^2\bigr).
\end{align}
\end{subequations}
Moreover, $\tens{F}^{-\top}=(\tens{V}\tens{R})^{-\top}=\tens{V}^{-1}\tens{R}$, whence
\begin{align}
\label{eq:P_SR_square}
&\tens{P}=\mathcal{J}\tens{T}\tens{V}^{-1}\tens{R}=\tens{M}\tens{R}, &&\text{where }\tens{M}=\mathcal{J}\tens{T}\tens{V}^{-1}.
\end{align}
Since $\tens{P}=\tau\tens{I}$, $\tens{M}=\tau\tens{R}^{-1}$ is a scalar multiple of a rotation. In components, this implies
\begin{align}
\label{eq:S_constraints_square}
M_{11} = M_{22}, \qquad  M_{12} = -M_{21}.
\end{align}
Computing the components of $\tens{M}$ using Eqs.~\eqref{eq:T_general_uniform_square} and simplifying constraints \eqref{eq:S_constraints_square} using \textsc{Mathematica} (Wolfram, Inc.) yields
\begin{subequations}\label{eq:factorized_constraints_square}
\begin{align}
(\lambda_1-\lambda_2)\left[-\kappa(\mathcal{J}-1)+\frac{\mu}{2\mathcal{J}^{2}}(\lambda_1+\lambda_2)^2\right]&=0, \\
(\lambda_1-\lambda_2)\left[A(\mathcal{J}-1)+\frac{K}{2\mathcal{J}^{2}}(\lambda_1+\lambda_2)^2\right]&=0.
\end{align}
\end{subequations}
The trivial solution $\lambda_1=\lambda_2$ is the square branch obtained above. The vanishing of the terms in square brackets defines additional solutions with broken symmetry. This yields a system of two linear equations in $\mathcal{J}-1$ and $(\lambda_1+\lambda_2)^2/2\mathcal{J}^2$, viz.,
\begin{subequations}\label{eq:branch_cond}
\begin{align}
\kappa(\mathcal{J}-1)-\mu\dfrac{(\lambda_1+\lambda_2)^2}{2\mathcal{J}^2}&=0,\label{eq:branch_cond1} \\
A(\mathcal{J}-1)+K\frac{(\lambda_1+\lambda_2)^2}{2\mathcal{J}^{2}}&=0.
\end{align}
\end{subequations}
Since $(\lambda_1+\lambda_2)^2/2\mathcal{J}^2>0$, this system must have a non-trivial solution, which happens if and only if
\begin{align}
\label{eq:para_match}
A\mu+\kappa K=0.
\end{align}
This is the parameter-matching condition required for a symmetry-breaking of the homogeneous branch discussed in the main text. 

Imposing Eqs.~\eqref{eq:branch_cond}, the expression for $\tens{M}=\mathcal{J}\tens{T}\tens{V}^{-1}$ simplifies to
\begin{align}
\tens{M}=\frac{\lambda_1+\lambda_2}{\mathcal{J}}\bigl(\mu \tens{I}+K\teps\bigr).\label{eq:M}
\end{align}
Let $\tens{R}=\mathcalbf{R}(\vartheta_1)$. Now $\tens{M}=\tau\tens{R}^{-1}=\tau\mathcalbf{R}(-\vartheta_1)$, so
\begin{align}
\tan{\vartheta_1}=-\dfrac{M_{12}}{M_{11}}=-\dfrac{K}{\mu}.
\end{align}
This exposes the geometric meaning of the parameter-matching condition~\eqref{eq:para_match}: 
\begin{align}
\tan{\vartheta_0}=\dfrac{A}{\kappa}=-\dfrac{K}{\mu}=\tan{\vartheta_1}.\label{eq:tant0t1}
\end{align}
In other words, the symmetry-breaking of the square branch requires the rotation of that branch to match the rotation of the deformations with broken symmetry.

Imposing this matching condition, $\mu\tens{I}+K\teps=\tens{R}^{-1}\sqrt{\mu^2+K^2}$. Using Eq.~\eqref{eq:M}, we then obtain
\begin{align}
\label{eq:P_identity_square}
\tau\tens{I}=\tens{P} =\tens{M}\tens{R} &=\frac{\lambda_1+\lambda_2}{\mathcal{J}}\bigl(\mu \tens{I}+K\teps\bigr)\tens{R} \nonumber\\
&=\sqrt{\mu^2+K^2}\,\frac{\lambda_1+\lambda_2}{\mathcal{J}}\,\tens{I}.
\end{align}
Hence, on the asymmetric branch,
\begin{align}
\label{eq:tau_sb_square}
\tau=\sqrt{\mu^2+K^2}\,\frac{\lambda_1+\lambda_2}{\mathcal{J}}.
\end{align}
Eliminating $\lambda_1+\lambda_2$ using Eq.~\eqref{eq:branch_cond1} yields
\begin{align}
\label{eq:J_of_tau_square}
\mathcal{J}=1+\frac{\mu\tau^2}{2\kappa(\mu^2+K^2)}.
\end{align}
Substituting back into Eq.~\eqref{eq:tau_sb_square}, we get
\begin{align}
\label{eq:s_of_tau_square}
\lambda_1+\lambda_2=\frac{\tau\,\mathcal{J}}{\sqrt{\mu^2+K^2}}=\frac{\tau}{\sqrt{\mu^2+K^2}}\left[1+\frac{\mu\tau^2}{2\kappa(\mu^2+K^2)}\right].
\end{align}
From definition \eqref{eq:F_general_uniform_square}, $\mathcal{J}=\lambda_1\lambda_2$. Hence $\lambda_1,\lambda_2$ satisfy the quadratic equation
\begin{align}
\lambda^2-\dfrac{\tau}{\sqrt{\mu^2+K^2}}\left[1\!+\!\frac{\mu\tau^2}{2\kappa(\mu^2\!+\!K^2)}\right]\lambda+\left[1\!+\!\frac{\mu\tau^2}{2\kappa(\mu^2\!+\!K^2)}\right]=0,
\end{align}
which parametrises the asymmetric branch of deformations. The square branch bifurcates to this asymmetric branch when $\lambda_1=\lambda_2=\lambda_\ast$. As $\lambda_\ast>0$, Eq.~\eqref{eq:s_of_tau_square} yields
\begin{align}
\tau_\ast=\dfrac{2}{\lambda_\ast}\sqrt{\mu^2+K^2}.\label{eq:tauast}
\end{align}
Equation~\eqref{eq:J_of_tau_square} then simplifies to
\begin{subequations}
\begin{align}
\lambda_\ast^2=1+\dfrac{2\mu}{\kappa\lambda_\ast^2},
\end{align}
a biquadratic equation with positive solution
\begin{align}
\lambda_\ast^2=\dfrac{1+\sqrt{1+8\mu/\kappa}}{2},
\end{align}
\end{subequations}
which is Eq.~(6) of the main text. Equation~\eqref{eq:tauast} finally gives the corresponding critical dead load.
\subsection{Shapes of the Rivlin square}
To describe the geometric shapes resulting from the deformations, we note that the edge vectors $\vec{e_1},\vec{e_2}$ of the undeformed square map to
\begin{align}
&\tens{F}\vec{e_1} =
\begin{pmatrix}
\lambda_1 \cos{\vartheta_0}\\
\lambda_2 \sin{\vartheta_0}
\end{pmatrix},&&\tens{F}\vec{e_2} =\begin{pmatrix}
-\lambda_1 \sin{\vartheta_0}\\
\lambda_2 \cos{\vartheta_0}
\end{pmatrix}.
\end{align}
These vectors define, in general, a parallelogram. Two special cases arise:
\begin{enumerate}[label=(\alph{enumi}),widest=2,leftmargin=*,itemsep=0pt,topsep=3pt]
\item The deformed shape is a rhombus if and only if ${\|\tens{F}\vec{e_1}\|=\|\tens{F}\vec{e_2}\|}$, which is equivalent to
\begin{align}
\bigl(\lambda_1^2-\lambda_2^2\bigr)\cos{2\vartheta_0}=0.
\end{align}
This holds on the symmetric branch $\lambda_1=\lambda_2$ and if $\cos{2\vartheta_0}=0\Leftrightarrow\left|\tan{\vartheta_0}\right|=1$. From Eq.~\eqref{eq:tant0t1} and assuming $\kappa,\mu>0$, the latter requires $A=\kappa$ and $K=-\mu$ or $A=-\kappa$ and $K=\mu$.
\item The deformed shape is a rectangle if and only if ${(\tens{F}\vec{e_1})\cdot(\tens{F}\vec{e_2})=0}$, which is if and only if
\begin{align}
\bigl(\lambda_1^2-\lambda_2^2\bigr)\sin{2\vartheta_0}=0.
\end{align}
Again, this holds on the symmetric branch $\lambda_1=\lambda_2$ and if $\sin{2\vartheta_0}=0$. For $\kappa,\mu>0$, $\tan{\vartheta_0}$ is well-defined from Eq.~\eqref{eq:tant0t1}, so the second possibility implies $\tan{\vartheta_0}=0$ and hence $A=K=0$. This is the even Rivlin square, as illustrated in Fig.~1(a) of the main text.
\end{enumerate}
In particular, both conditions hold on the symmetric branch, on which the shapes are therefore squares, as expected.

\subsection{Inhomogeneous deformations of the Rivlin square}
As announced in the main text, we analyse inhomogeneous deformations by introducing an infinitesimal displacement
\begin{align}
\delta\vec{u}(\vec{X})=\vec{a}\,\e^{\i \vec{q}\cdot\vec{X}},
\end{align}
in which $\vec{q}$ is a polarisation vector and $\vec{a}$ is the displacement magnitude. We solve the corresponding equations of incremental elasticity~\cite{ogden} about the homogeneous base state $\tens{F}_\tzero=\lambda\tens{R}=\lambda\mathcalbf{R}(\vartheta_0)$, from Eqs.~\eqref{eq:Fsym}, in which $\lambda$ is given by the second of Eqs.~\eqref{eq:symsol}. The resulting deformation gradient is $\tens{F}=\tens{F}_\tzero+\delta\tens{F}$, where
\begin{align}
\delta\tens{F}=\operatorname{Grad}{\delta\vec{u}}=\i\,\vec{a}\otimes\vec{q}\,\e^{\i\vec{q}\cdot\vec{X}}.\label{eq:delta_F}
\end{align}
In what follows, we will drop the factor $\e^{\i\vec{q}\cdot\vec{X}}$ implicitly. We write $\tens{R}\vec{q}=q\vec{m}$, where $q=\|\vec{q}\|$, so that $\|\vec{m}\|=1$ because $\tens{R}$ is a rotation. We let $\vec{t}=\teps\vec{m}$, so $\{\vec{m},\vec{t}\}$ is orthonormal.

Now, in the homogeneous base state, $\mathcal{J}_0=\lambda^2$, so, from Jacobi's formula~\cite{goriely},
\begin{subequations}\label{eq:deltaeverything}
\begin{align}
\delta\mathcal{J}=\mathcal{J}_0\,\tr{\bigl(\tens{F}_\tzero^{-1}\delta\tens{F}\bigr)}&=\lambda^2\,\tr{\big(\i\lambda^{-1}\tens{R}^\top(\vec{a}\otimes\vec{q})\big)}\nonumber\\
&=\i\lambda\,\vec{a}\cdot(\tens{R}\vec{q}) = \i \lambda q a_m, \label{eq:dJ}
\end{align}
where we have decomposed $\vec{a}=a_m\vec{m}+a_t\vec{t}$, so that $a_m=\vec{a}\cdot\vec{m}$, $a_t=\vec{a}\cdot\vec{t}$. Moreover,
\begin{align}
\delta\tens{B} &=\delta\tens{F}\,\tens{F}_\tzero^\top+\tens{F}_\tzero\,\delta\tens{F}^\top =\i\lambda q\big(\vec{a}\otimes\vec{m}+\vec{m}\otimes\vec{a}\big).\label{eq:dB}
\end{align}
Hence
\begin{align}
\tr{\delta\tens{B}}&=2\i\lambda q a_m, \\
\delta\dev{\tens{B}}\,\vec{m}&=\dev{\delta \tens{B}}\,\vec{m}=\left(\delta\tens{B}-\dfrac{\tr{\delta\tens{B}}}{2}\tens{I}\right)\vec{m}\nonumber\\
&=\i\lambda q\left(\vec{a}\otimes\vec{m}+\vec{m}\otimes\vec{a}-a_m\tens{I}\right)\vec{m}\nonumber\\
&=\i\lambda q\,\vec{a}.
\end{align}
Moreover, using $\teps\vec{m}=\vec{t}$, $\teps\vec{t}=-\vec{m}$ and Eq.~\eqref{eq:dB} again,
\begin{align}
\delta[\tens{B},\teps]\,\vec{m}&=[\delta\tens{B},\teps]\,\vec{m}=\delta\tens{B}(\teps\vec{m})-\teps(\delta\tens{B}\,\vec{m})\nonumber\\
&=\delta\tens{B}\,\vec{t}-\i\lambda q\teps\big(\vec{a}+a_m\vec{m}\big)\nonumber\\
&=\i\lambda q[a_t\vec{m}-\teps(2a_m\vec{m}+a_t\vec{t})]\nonumber\\
&=\i\lambda q[a_t\vec{m}-(2a_m\vec{t}-a_t\vec{m})]\nonumber\\
&=2\i\lambda q(a_t\vec{m}-a_m\vec{t}).\label{eq:dBem}
\end{align}
\end{subequations}
Let $\tens{T}_\tzero$ be the Cauchy stress at the isotropic base state $\tens{F}_\tzero$, and let $\delta\tens{T}$ be the corresponding increment. Linearising Eq.~\eqref{eq:comp_T} yields
\begin{align}
\delta\tens{T}=\kappa\,\delta\mathcal{J}\,\tens{I}-A\,\delta\mathcal{J}\,\teps+\mu\lambda^{-4}\delta\dev{\tens{B}}-\frac{K}{2}\lambda^{-4}\delta[\tens{B},\teps].\label{eq:delta_T}
\end{align}
Using Eqs.~\eqref{eq:deltaeverything}, we obtain
\begin{align}
\delta\tens{T}\,\vec{m}&=\kappa\delta\mathcal{J}\,\vec{m}-A\delta\mathcal{J}\teps\vec{m}+\mu\lambda^{-4}\delta\dev{\tens{B}}\vec{m}\nonumber\\
&\qquad-\frac{K}{2}\lambda^{-4}\delta[\tens{B},\teps]\vec{m}\nonumber\\
&=\kappa(\i\lambda q a_m)\vec{m}-A(\i\lambda q a_m)\vec{t}+\mu\lambda^{-4}(\i\lambda q\,\vec{a})\nonumber\\
&\qquad-\frac{K}{2}\lambda^{-4}\bigl[2\i\lambda q(a_t\vec{m}-a_m\vec{t})\bigr]\nonumber\\
&=\i q\bigl[\bigl(\kappa\lambda+\mu\lambda^{-3}\bigr)a_m-K\lambda^{-3}a_t\bigr]\vec{m}\nonumber\\
&\qquad+\i q\bigl[\bigl(-A\lambda+K\lambda^{-3}\bigr)a_m+\mu\lambda^{-3}a_t\bigr]\vec{t}.\label{eq:deltaTm}
\end{align}
The first Piola--Kirchhoff stress at $\tens{F}_\tzero$ is $\smash{\tens{P}_\tzero=\mathcal{J}_0\tens{T}_\tzero\tens{F}_{\smash{\tzero}}^{-\top}}$, with increment $\delta\tens{P}$. By definition $\tens{P}=\mathcal{J}\tens{T}\tens{F}^{-\top}$, so
\begin{align}
\delta\tens{P}=\delta\mathcal{J}\,\tens{T}_\tzero\tens{F}_{\smash{\tzero}}^{-\top}+\mathcal{J}_0\,\delta\tens{T}\,\tens{F}_{\smash{\tzero}}^{-\top}+\mathcal{J}_0\,\tens{T}_\tzero\,\delta\bigl(\tens{F}^{-\top}\bigr).\label{eq:delta_P}
\end{align}
Now
\begin{subequations}
\begin{align}
\tens{F}^{-\top}&=(\tens{F}_\tzero+\delta\tens{F})^{-\top}=\bigl(\tens{F}_\tzero\bigl(\tens{I}+\tens{F}_{\smash{\tzero}}^{-1}\delta\tens{F}\bigr)\bigr)^{-\top}\nonumber\\
&=\tens{F}_{\smash{\tzero}}^{-\top}\bigl(\tens{I}-\tens{F}_{\smash{\tzero}}^{-1}\delta\tens{F}\bigr)^\top+\mathcal{O}\bigl(\|\delta\tens{F}\|^2\bigr)\nonumber\\
&=\tens{F}_{\smash{\tzero}}^{-\top}-\tens{F}_{\smash{\tzero}}^{-\top}\delta\tens{F}\,\tens{F}_{\smash{\tzero}}^{-\top}+\mathcal{O}\bigl(\|\delta\tens{F}\|^2\bigr),
\end{align}
so
\begin{align}
\delta\bigl(\tens{F}^{-\top}\bigr)=-\tens{F}_{\smash{\tzero}}^{-\top}\delta\tens{F}\,\tens{F}_{\smash{\tzero}}^{-\top}.\label{eq:deltaFmT}
\end{align}
\end{subequations}
The incremental force balance equation~\cite{ogden} is $\Div{\delta\tens{P}}=\vec{0}$. For our plane-wave ansatz, $\Div{\delta\tens{P}}=\i\,\delta\tens{P}\,\vec{q}$. We now compute $\tens{F}_{\smash{\tzero}}^{-\top}=\lambda^{-1}\tens{R}$ with $\tens{R}\vec{q}=q\vec{m}$ by definition, so
\begin{subequations}\label{eq:deltaFq}
\begin{align}
\tens{F}_\tzero^{-\top}\vec{q}=\lambda^{-1}\tens{R}\vec{q}=\lambda^{-1}q\vec{m}.\label{eq:F0mTq}
\end{align}
Moreover, since $\delta\tens{F}^\top=\i\,\vec{q}\otimes\vec{a}$ from Eq.~\eqref{eq:delta_F}, Eqs.~\eqref{eq:deltaFmT} and~\eqref{eq:F0mTq} give
\begin{align}
\delta\bigl(\tens{F}^{-\top}\bigr)\,\vec{q}&=-\tens{F}_{\smash{\tzero}}^{-\top}\delta\tens{F}^\top\tens{F}_{\smash{\tzero}}^{-\top}\vec{q}=-\lambda^{-1}\tens{R}(\i\,\vec{q}\otimes\vec{a})\bigl(\lambda^{-1}q\vec{m}\bigr)\nonumber\\
&=-\i\lambda^{-2}q(\vec{a}\cdot\vec{m})\tens{R}\vec{q}=-\i\lambda^{-2}q^2 a_m\,\vec{m},
\end{align}
\end{subequations}
using $\tens{R}\vec{q}=q\vec{m}$ again. We now get an expression for $\delta\tens{P}\,\vec{q}$ from Eq.~\eqref{eq:delta_P}, in which the first and third terms cancel because
\begin{align}
&\delta\mathcal{J}\,\tens{T}_\tzero\tens{F}_\tzero^{-\top}\vec{q}+\mathcal{J}_0\,\tens{T}_\tzero\,\delta(\tens{F}^{-\top})\,\vec{q}\nonumber\\
&=(\i\lambda q a_m)\tens{T}_\tzero(\lambda^{-1}q\vec{m})+\lambda^2\tens{T}_\tzero\bigl(-\i\lambda^{-2}q^2 a_m\vec{m}\bigr)=\vec{0},
\end{align}
using Eqs.~\eqref{eq:dJ} and~\eqref{eq:deltaFq}. Hence, from Eq.~\eqref{eq:deltaTm},
\begin{align}
\delta\tens{P}\,\vec{q}&=\mathcal{J}_0\,\delta\tens{T}\,\tens{F}_\tzero^{-\top}\vec{q}=\lambda q\,\delta\tens{T}\,\vec{m}\nonumber\\
&=\i q^2\Bigl[\bigl(\kappa\lambda^2+\mu\lambda^{-2}\bigr)a_m-K\lambda^{-2}a_t\Bigr]\vec{m}\nonumber\\
&\qquad+\i q^2\Bigl[\bigl(-A\lambda^2+K\lambda^{-2}\bigr)a_m+\mu\lambda^{-2}a_t\Bigr]\vec{t}.
\label{eq:dP_q}
\end{align}
Writing $\delta\tens{P}\,\vec{q}=\i q^2\tens{Q}(\vec{q})\vec{a}$ defines the acoustic tensor~\cite{ogden}. Explicitly, with respect to the natural basis $\{\vec{m},\vec{t}\}$,
\begin{align}
\tens{Q}(\vec{q})=\begin{pmatrix}
\kappa\lambda^2+\mu\lambda^{-2} & -K\lambda^{-2}\\
K\lambda^{-2}-A\lambda^2 & \mu\lambda^{-2}
\end{pmatrix}.
\end{align}
The incremental force balance equation is now $\tens{Q}(\vec{q})\vec{a}=\vec{0}$, which has a nontrivial solution $\vec{a}\neq\vec{0}$ if and only if
\begin{align}
0=\det\tens{Q}(\vec{q})&=\kappa\mu-AK+\frac{\mu^2+K^2}{\lambda^4}. \label{eq:detQ}
\end{align}
This is independent of $\vec{q}$, as expected because the base state is isotropic. As discussed in the main text, this requires 
\begin{align}
AK>\kappa\mu.
\label{eq:para_cond}
\end{align}
This condition is, for positive shear and bulk moduli, $\kappa,\mu>0$, incompatible with the homogeneous bifurcation branch obtained in the previous subsection, as raised in the main text, too. If condition~\eqref{eq:para_cond} holds, then inhomogeneous solutions become possible at $\lambda=\lambda_{\ast\ast}$, where, from Eq.~\eqref{eq:detQ},
\begin{align}
\lambda_{\ast\ast}^4= \dfrac{\mu^2+K^2}{AK-\kappa\mu},
\end{align}
which is Eq.~(8) of the main text. Let $r=a_t/a_m$ be the polarisation ratio. At $\lambda=\lambda_{\ast\ast}$, $\tens{Q}(\vec{q})\vec{a}=\vec{0}$ requires
\begin{align}
\bigl(\kappa\lambda_{\ast\ast}^2+\mu\lambda_{\ast\ast}^{-2}\bigr)-K\lambda_{\ast\ast}^{-2}r=\bigl(K\lambda_{\ast\ast}^{-2}-A\lambda_{\ast\ast}^2\bigr)+\mu\lambda_{\ast\ast}^{-2}r=0,\label{eq:critids}
\end{align}
which implies
\begin{align}
r=\frac{\kappa\lambda_{\ast\ast}^4+\mu}{K}=\frac{A\lambda_{\ast\ast}^4-K}{\mu}.\label{eq:polar_ratio}
\end{align}
Again, these results are independent of $q$, so all wavelengths are simultaneously marginal. Any selection mechanism of a finite wavelength therefore requires some additional regularisation or nonlinearity. 
\subsubsection*{\textbf{Critical deformation modes}}
The single critical modes do not in general satisfy the incremental boundary conditions $\delta\tens{P}\cdot\vec{N}=\vec{0}$, where $\vec{N}$ is the normal to the sides of the square $\mathcal{S}=[0,1]\times[0,1]$.

At the critical stretch $\lambda=\lambda_{\ast\ast}$, Eq.~\eqref{eq:polar_ratio} determines the polarisation ratio $r$. We recall that $\vec m=\tens{R}\vec q/q$ and ${\vec t=\teps\vec m=\teps\tens{R}\vec q/q}$, with $q=\|\vec q\|$, so the polarisation of each critical Fourier mode can be written as
\begin{align}
\vec a&=a_m\vec m+a_t\vec t=a_m(\vec m+r\vec t)=\frac{a_m}{q}\bigl(\tens{R}\vec q+r\,\teps\tens{R}\vec q\bigr).
\end{align}
Any critical displacement is a superposition of critical modes of this form, viz., 
\begin{subequations}
\begin{align}
\delta\vec{u}(\vec{X})=\sum_{n=1}^{\infty}c_n\bigl(\tens{R}+r\,\teps\tens{R}\bigr)\vec{q}_n\e^{\i\vec{q}_n\cdot\vec{X}},
\end{align}
for suitable complex coefficients $c_n$. Equivalently, noting that $\vec\nabla\e^{\i\vec{q}_n\cdot\vec{X}}=\i\vec{q}_n\e^{\i\vec{q}_n\cdot\vec{X}}$, we may write
\begin{align}
&\delta\vec{u}(\vec{X})=\bigl(\tens{R}+r\,\teps\tens{R}\bigr)\vec\nabla\phi(\vec{X}), &&\text{with }\phi(\vec{X})=-\i\sum_{n=1}^{\infty}c_n\,\e^{\i\vec{q}_n\cdot\vec{X}}. \label{eq:u_phi_square}
\end{align}
\end{subequations}
Any such displacement $\delta\vec{u}(\vec{X})$ satisfies the incremental equation $\Div{\delta\tens{P}}=\vec{0}$ by construction. To solve the problem on the square $\mathcal{S}$, it therefore suffices to satisfy the exact boundary condition $\delta\tens{P}\cdot\vec{N}=\vec{0}$ on its boundary $\partial\mathcal{S}$. Now, from Eq.~\eqref{eq:u_phi_square}, 
\begin{align}
\delta\tens{F}=\vec\nabla\delta\vec{u}=\bigl(\tens{R}+r\,\teps\tens{R}\bigr)\vabla\vabla\phi=\tens{R}(\tens{I}+r\teps)\vabla\vabla\phi,
\label{eq:dF_phi_square}
\end{align}
noting that $\tens{R}^\top\teps\tens{R}=\teps$. Recalling that $\tens{F}_\tzero^{-1}=\lambda^{-1}\tens{R}^\top$, $\mathcal{J}_0=\lambda^2$, and using the symmetry of $\vabla\vabla\phi$, Jacobi's formula~\cite{goriely} gives
\begin{align}
\delta\mathcal{J}&=\mathcal{J}_0\,\tr\bigl(\tens{F}_\tzero^{-1}\delta\tens{F}\bigr)=\lambda^2\,\tr\left[\lambda^{-1}\tens{R}^\top(\tens{R}+r\,\teps\tens{R})\vabla\vabla\phi\right]\nonumber\\
&=\lambda\,\tr\left[(\tens{I}+r\,\teps)\vabla\vabla\phi\right]=\lambda\,\Delta\phi,\label{eq:dJ_phi_square}
\end{align}
where $\Delta$ is the Laplacian. Taking the transpose of Eq.~\eqref{eq:dF_phi_square}, we obtain
\begin{align}
\delta\tens{F}^\top&=(\vabla\vabla\phi)\,\left(\tens{I}-r\,\teps\right)\tens{R}^\top.
\end{align}
Hence, using Eq.~\eqref{eq:deltaFmT},
\begin{align}
\delta\bigl(\tens{F}^{-\top}\bigr)&=-\tens{F}_\tzero^{-\top}\delta\tens{F}^{\top}\tens{F}_\tzero^{-\top}=-\lambda^{-2}\tens{R}\,(\vabla\vabla\phi)\,(\tens{I}-r\,\teps)\tens{R}^\top\tens{R}\nonumber\\
&=-\lambda^{-2}\tens{R}\,(\vabla\vabla\phi)\,(\tens{I}-r\,\teps). \label{eq:dFinvT_phi_square}
\end{align}
Next,
\begin{align}
\delta\tens{B}&=\delta\tens{F}\,\tens{F}_\tzero^\top+\tens{F}_\tzero\,\delta\tens{F}^\top\nonumber\\
&=\tens{R}(\tens{I}+r\,\teps)(\vabla\vabla\phi)\,\lambda\tens{R}^\top+\lambda\tens{R}\,(\vabla\vabla\phi)\,(\tens{I}-r\,\teps)\tens{R}^\top\nonumber\\
&=\lambda\tens{R}\bigl[(\tens{I}+r\,\teps)(\vabla\vabla\phi)+(\vabla\vabla\phi)\,(\tens{I}-r\,\teps)\bigr]\tens{R}^\top\nonumber\\
&=\lambda\tens{R}\Bigl(2\vabla\vabla\phi+r[\teps,\vabla\vabla\phi]\Bigr)\tens{R}^\top. \label{eq:dB_phi_square}
\end{align}
To simplify the final term, we compute
\begin{align}
[\teps,\vabla\vabla\phi]=\begin{pmatrix}-2\phi_{,12} & \phi_{,11}-\phi_{,22}\\ \phi_{,11}-\phi_{,22} & 2\phi_{,12}\end{pmatrix}. \label{eq:epsHminusHeps_square}
\end{align}
This shows that $\tr{[\teps,\vabla\vabla\phi]}=0$, so Eq.~\eqref{eq:dB_phi_square} yields
\begin{align}
\tr{\delta\tens{B}}&=\lambda\,\tr{(2\vabla\vabla\phi)}=2\lambda\,\Delta\phi=2\lambda\tens{R}(\Delta\phi\,\tens{I})\tens{R}^\top.
\end{align}
Hence
\begin{align}
\delta\dev{\tens{B}}&=\delta\tens{B}-\frac{\tr(\delta\tens{B})}{2}\tens{I}\nonumber\\
&=\lambda\tens{R}\bigl[2(\vabla\vabla\phi)-(\Delta\phi)\tens{I}+r[\teps,\vabla\vabla\phi]\bigr]\tens{R}^\top. \label{eq:ddevB_phi_square}
\end{align}
Equation~\eqref{eq:dB_phi_square} also yields
\begin{align}
\delta[\tens{B},\teps]&=\lambda\bigl[\tens{R}\bigl[2(\vabla\vabla\phi)+r[\teps,\vabla\vabla\phi]\bigr]\tens{R}^\top,\teps\bigr]\nonumber\\
&=\lambda\tens{R}\bigl[2(\vabla\vabla\phi)+r[\teps,\vabla\vabla\phi],\teps\bigr]\tens{R}^\top\nonumber\\
&=\lambda\tens{R}\bigl(2[\vabla\vabla\phi,\teps]+r\bigl[[\teps,\vabla\vabla\phi],\teps\bigr]\bigr)\tens{R}^\top. \label{eq:dBeps_phi_square}
\end{align}
From Eq.~\eqref{eq:epsHminusHeps_square} we compute, explicitly, 
\begin{align}
\bigl[[\teps,\vabla\vabla\phi],\teps\bigr]
&=2\begin{pmatrix}\phi_{,11}-\phi_{,22} & 2\phi_{,12}\\ 2\phi_{,12} & \phi_{,22}-\phi_{,11}\end{pmatrix}\nonumber\\
&=2\left[2(\vabla\vabla\phi)-(\Delta\phi)\tens{I}\right]. \label{eq:doublecomm_square}
\end{align}
Substituting Eqs.~\eqref{eq:dJ_phi_square}, \eqref{eq:ddevB_phi_square}, \eqref{eq:dBeps_phi_square}, and \eqref{eq:doublecomm_square} into Eq.~\eqref{eq:delta_T}, we obtain
\begin{subequations}
\begin{align}
\lambda\,\delta\tens{T}&=\kappa\lambda^2(\Delta\phi)\tens{I}-A\lambda^2(\Delta\phi)\teps\nonumber\\
&\qquad+\mu\lambda^{-2}\tens{R}\bigl[2(\vabla\vabla\phi)-(\Delta\phi)\tens{I}+r[\teps,\vabla\vabla\phi]\bigr]\tens{R}^\top\nonumber\\
&\qquad-K\lambda^{-2}\tens{R}\bigl([\vabla\vabla\phi,\teps]+r[2(\vabla\vabla\phi)-(\Delta\phi)\tens{I}]\bigr)\tens{R}^\top\nonumber\\
&=\tens{R}\bigl\{\kappa\lambda^2(\Delta\phi)\tens{I}-A\lambda^2(\Delta\phi)\teps\nonumber\\
&\qquad+\lambda^{-2}(\mu-Kr)\left[2(\vabla\vabla\phi)-(\Delta\phi)\tens{I}\right]\nonumber\\
&\qquad+\lambda^{-2}(\mu r+K)[\teps,\vabla\vabla\phi]\bigr\}\tens{R}^\top. \label{eq:dT_expand3_square}
\end{align}
Now $\mu-Kr=-\kappa\lambda^4$, $\mu r+K=A\lambda^4$ from Eqs.~\eqref{eq:critids}, so this becomes
\begin{align}
\lambda\,\delta\tens{T}&=\lambda^2\tens{R}\bigl\{\kappa(\Delta\phi)\tens{I}-A(\Delta\phi)\teps-\kappa[2\vabla\vabla\phi-(\Delta\phi)\tens{I}]\nonumber\\
&\qquad\qquad+A[\teps,\vabla\vabla\phi]\bigr\}\tens{R}^\top\nonumber\\
&=\lambda^2\tens{R}\bigl\{2\kappa\bigl[(\Delta\phi)\tens{I}-\vabla\vabla\phi\bigr]\nonumber\\
&\qquad\qquad+A\bigl[[\teps,\vabla\vabla\phi]-(\Delta\phi)\teps\bigr]\bigr\}\tens{R}^\top. \label{eq:dT_expand4_square}
\end{align}
\end{subequations}
At this stage, it is convenient to introduce the matrix of cofactors,
\begin{align}
\operatorname{cof}\bigl(\vabla\vabla\phi\bigr)=\begin{pmatrix}\phi_{,22} & -\phi_{,12}\\ -\phi_{,12} & \phi_{,11}\end{pmatrix}=(\Delta\phi)\tens{I}-\vabla\vabla\phi. \label{eq:cofH_square}
\end{align}
We now compute
\begin{align}
[\teps,\vabla\vabla\phi]-(\Delta\phi)\teps&=\begin{pmatrix}-2\phi_{,12} & 2\phi_{,11}\\ -2\phi_{,22} & 2\phi_{,12}\end{pmatrix}=-2\teps\,\operatorname{cof}\bigl(\vabla\vabla\phi\bigr). \label{eq:epscof_square}
\end{align}
Substituting Eqs.~\eqref{eq:cofH_square} and \eqref{eq:epscof_square} into Eq.~\eqref{eq:dT_expand4_square}, we finally obtain
\begin{align}
\lambda\,\delta\tens{T}&=2\lambda^2\tens{R}\bigl[\kappa\operatorname{cof}\bigl(\vabla\vabla\phi\bigr)-A\teps\operatorname{cof}\bigl(\vabla\vabla\phi\bigr)\bigr]\tens{R}^\top\nonumber\\
&=2\lambda^2\tens{R}(\kappa\tens{I}-A\teps)\operatorname{cof}{\bigl(\vabla\vabla\phi\bigr)}\tens{R}^\top. \label{eq:dT_phi_square}
\end{align}
As already noted above Eq.~\eqref{eq:Psq_pre}, the Cauchy stress in the isotropic base state is
\begin{align}
\tens{T}_\tzero=\kappa(\lambda^2-1)\tens{I}-A(\lambda^2-1)\teps=(\lambda^2-1)(\kappa\tens{I}-A\teps). \label{eq:T0_square}
\end{align}
From Eq.~\eqref{eq:delta_P}, and using Eqs.~\eqref{eq:dJ_phi_square}, and \eqref{eq:dFinvT_phi_square}, together with $\tens{F}_{\smash{\tzero}}^{-\top}=\lambda^{-1}\tens{R}$, $\mathcal{J}_0=\lambda^2$, we obtain
\begin{align}
\delta\tens{P}&=\delta\mathcal{J}\,\tens{T}_\tzero\tens{F}_\tzero^{-\top}+\mathcal{J}_0\,\delta\tens{T}\,\tens{F}_\tzero^{-\top}+\mathcal{J}_0\,\tens{T}_\tzero\,\delta\bigl(\tens{F}^{-\top}\bigr)\nonumber\\
&=\lambda(\Delta\phi)\,\tens{T}_\tzero(\lambda^{-1}\tens{R})+\lambda^2\delta\tens{T}(\lambda^{-1}\tens{R})\nonumber\\
&\qquad+\lambda^2\tens{T}_\tzero\bigl[-\lambda^{-2}\tens{R}(\vabla\vabla\phi)\,(\tens{I}-r\,\teps)\bigr]\nonumber\\
&=(\Delta\phi)\,\tens{T}_\tzero\tens{R}+\lambda\,\delta\tens{T}\,\tens{R}-\tens{T}_\tzero\tens{R}\,(\vabla\vabla\phi)\,(\tens{I}-r\,\teps). \label{eq:dP_expand1_square}
\end{align}
Now using Eqs.~\eqref{eq:dT_phi_square} and~\eqref{eq:T0_square} and since $\tens{R}$ commutes with $\teps$ and hence with $\tens{T}_\tzero$ from Eq.~\eqref{eq:T0_square},
\begin{subequations}
\begin{align}
(\Delta\phi)\,\tens{T}_\tzero\tens{R}&=\tens{R}(\kappa\tens{I}-A\,\teps)(\lambda^2-1)(\Delta\phi)\tens{I},\\
\lambda\,\delta\tens{T}\,\tens{R}
&=2\lambda^2\tens{R}(\kappa\tens{I}-A\,\teps)\operatorname{cof}\bigl(\vabla\vabla\phi\bigr),
\end{align}
and
\begin{align}
&\tens{T}_\tzero\tens{R}\,(\vabla\vabla\phi)\,(\tens{I}-r\,\teps)\nonumber\\
&\quad=\tens{R}(\kappa\tens{I}-A\,\teps)(\lambda^2-1)(\vabla\vabla\phi)\,(\tens{I}-r\,\teps).
\end{align}
\end{subequations}
Substituting these three expressions into Eq.~\eqref{eq:dP_expand1_square} and factoring out the common left factor $\tens{R}(\kappa\tens{I}-A\,\teps)$, we get
\begin{align}
\delta\tens{P}&=\tens{R}(\kappa\tens{I}-A\,\teps)\bigl[(\lambda^2-1)(\Delta\phi)\tens{I}+2\lambda^2\operatorname{cof}\bigl(\vabla\vabla\phi\bigr)\nonumber\\
&\qquad\qquad\qquad\qquad-(\lambda^2-1)\vabla\vabla\phi+(\lambda^2-1)r\,\vabla\vabla\phi\,\teps\bigr]. \label{eq:dP_expand3_square}
\end{align}
Using Eq.~\eqref{eq:cofH_square}, we observe that the first, second, and third terms within the square brackets combine as
\begin{align}
&(\lambda^2-1)(\Delta\phi)\tens{I}+2\lambda^2\operatorname{cof}\bigl(\vabla\vabla\phi\bigr)-(\lambda^2-1)\vabla\vabla\phi\nonumber\\
&\quad=(3\lambda^2-1)\operatorname{cof}\bigl(\vabla\vabla\phi\bigr).
\end{align}
Equation~\eqref{eq:dP_expand3_square} thus becomes
\begin{align}
\delta\tens{P}=\tens{R}(\kappa\tens{I}-A\,\teps)\bigl[(3\lambda^2\!-\!1)\operatorname{cof}\bigl(\vabla\vabla\phi\bigr)+(\lambda^2\!-\!1)r\,\vabla\vabla^2\phi\,\teps\bigr]. \label{eq:dP_expand4_square}
\end{align}
By direct computation, we also observe that
\begin{align}
\operatorname{cof}\bigl(\vabla\vabla\phi\bigr)=-\teps\,(\vabla\vabla\phi)\,\teps,
\end{align}
which yields, using $\teps^2=-\tens{I}$,
\begin{align}
\delta\tens{P}&=\tens{R}\bigl[(\lambda^2\!-\!1)r(\kappa\tens{I}-A\teps)-(3\lambda^2\!-\!1)(A\tens{I}+\kappa\teps)\bigr](\vabla\vabla\phi)\,\teps\nonumber\\&=\tens{R}\bigl(\eta_2\tens{I}-\eta_1\teps\bigr)(\vabla\vabla\phi)\,\teps, \label{eq:dP_phi_square}
\end{align}
where
\begin{subequations}
\begin{align}
\eta_1&=\kappa(3\lambda^2-1)+Ar(\lambda^2-1),\\
\eta_2&=\kappa r(\lambda^2-1)-A(3\lambda^2-1). \label{eq:c1c2_square}
\end{align}
\end{subequations}
This makes explicit the point noted earlier, that the incremental equilibrium equation holds identically in the bulk: now
\begin{align}
\Div{\delta\tens{P}}&=\tens{R}\bigl(\eta_2\tens{I}-\eta_1\teps\bigr)\Div{\bigl((\vabla\vabla\phi)\,\teps\bigr)} = \vec{0}, \label{eq:DivP_phi_square}
\end{align}
because $\smash{\Div{\bigl((\vabla\vabla\phi)\,\teps\bigr)}_i=\partial_j\bigl(\phi_{,ik}\varepsilon_{kj}\bigr)=\varepsilon_{kj}\phi_{,ijk}=0}$, since $\phi_{,ijk}$ is symmetric and $\varepsilon_{kj}$ is antisymmetric in $j,k$. 

Finally, we can impose the exact incremental boundary condition $\delta\tens{P}\cdot\vec{N}=\vec{0}$ on $\partial\mathcal{S}$. From Eq.~\eqref{eq:dP_phi_square}, we have
\begin{align}
\delta\tens{P}\cdot\vec N=\tens{R}\bigl(\eta_2\tens{I}-\eta_1\teps\bigr)\vabla\vabla\phi\,\teps\cdot\vec N. \label{eq:bc_general_square}
\end{align}
Since $\tens{R}$ is a rotation and $\det\bigl(\eta_2\tens{I}-\eta_1\teps\bigr)=\eta_1^2+\eta_2^2$, the prefactor $\tens{R}(\eta_2\tens{I}-\eta_1\teps)$ is invertible unless $\eta_1=\eta_2=0$. From Eqs.~\eqref{eq:c1c2_square}, this would define simultaneous linear equations for $3\lambda^2-1,r(\lambda^2-1)$, which only allow the trivial solution $3\lambda^2-1=r(\lambda^2-1)=0$ unless $A^2+\kappa^2=0$. The second condition cannot hold for we expect $\kappa>0$ as discussed repeatedly in the main text; the trivial solution implies $\lambda^2=1/3$ and hence $r=0$, which contradicts Eq.~\eqref{eq:polar_ratio} if $\kappa,\mu>0$. We may therefore assume that $\tens{R}(\eta_2\tens{I}-\eta_1\teps)$ is invertible. The boundary conditions now become
\begin{align}
(\vabla\vabla\phi)\,\teps\cdot\vec N=\vec 0 \qquad \text{on } \partial\mathcal{S}. \label{eq:bc_reduced_general_square}
\end{align}
These expand to
\begin{subequations}\label{eq:phi_bc_square}
\begin{align}
\phi_{,22}=\phi_{,12}=0 \quad \text{on } X_1=0,1, \\
\phi_{,11}=\phi_{,12}=0 \quad \text{on } X_2=0,1. 
\end{align}
\end{subequations}
In Fig.~2 of the main text, we plot the modes associated for a specific choice of $\phi$ that clearly satisfies Eqs.~\eqref{eq:phi_bc_square},
\begin{align}
\phi(X_1,X_2)=\sin^2{(m\pi X_1)}\sin^2{(n\pi X_2)}, \quad\text{for } m,n\in\mathbb{N}_+.
\end{align}

\section{\uppercase{The incompressible odd Rivlin square}}
We next consider the incompressible case. As in the main text, we take the incompressible Cauchy stress to be given by Eq.~\eqref{eq:incomp_T}.

\subsection{Homogeneous base state and perturbation}
We first identify the homogeneous isotropic solution. Since incompressibility requires $\det{\tens{F}}=1$, the isotropic branch is simply
\begin{align}
\tens{F}_\tzero=\tens{I},\qquad \tens{B}_\tzero=\tens{I},\qquad \mathcal{J}_0=1.
\end{align}
Because $[\tens{B}_\tzero,\teps]=[\tens{I},\teps]=\tzero$, the odd contribution to Eq.~\eqref{eq:incomp_T} vanishes in this base state and
\begin{align}
\tens{T}_\tzero=(-p_0+\mu)\tens{I}.
\end{align}
Since $\tens{F}_{\smash{\tzero}}^{-\top}=\tens{I}$ and $\mathcal{J}_0=1$, the first Piola--Kirchhoff stress is simply $\tens{P}_\tzero=\tens{T}_\tzero$. Imposing the dead-load condition $\tens{P}_\tzero=\tau\tens{I}$ gives
\begin{align}
&\tens{T}_\tzero=\tau\tens{I},&&p_0=\mu-\tau.
\end{align}
We now perturb about this base state by writing
\begin{align}
\tens{F}=\tens{F}_\tzero+\delta\tens{F},\qquad p=p_0+ \delta p.
\end{align}
In particular, the incremental incompressibility constraint is, from Jacobi's formula~\cite{goriely}, 
\begin{align}
\delta\mathcal{J}=\tr{\delta\tens{F}}=0.
\label{eq:incomp_u}
\end{align}
Moreover,
\begin{align}
\delta\tens{B}=\delta\tens{F}+\delta\tens{F}^\top.
\end{align}
From Eq.~\eqref{eq:incomp_T}, the increment of the Cauchy stress is
\begin{align}
\delta\tens{T}=-\delta p\,\tens{I}+\mu\,\delta\tens{B}-\frac{K}{2}\delta[\tens{B},\teps].
\end{align}
Since $\tens{F}_\tzero=\tens{I}$ and, from Eq.~\eqref{eq:deltaFmT}, $\delta\bigl(\tens{F}^{-\top}\bigr)=-\delta\tens{F}^\top$, so
\begin{align}
\delta\tens{P}&=\delta\tens{T}+\tens{T}_\tzero\,\delta(\tens{F}^{-\top})=\delta\tens{T}-\tau\,\delta\tens{F}^\top.
\end{align}
The increment of the first Piola--Kirchhoff stress is therefore
\begin{align}
\delta\tens{P}&=\mu\bigl(\delta\tens{F}+\delta\tens{F}^\top\bigr)-\frac{K}{2}\bigl[\bigl(\delta\tens{F}+\delta\tens{F}^\top\bigr)\teps-\teps\bigl(\delta\tens{F}+\delta\tens{F}^\top\bigr)\bigr]\nonumber\\
&\qquad-\tau\,\delta\tens{F}^\top-\delta p\,\tens{I}.
\label{eq:deltaP_final}
\end{align}
\subsection{Homogeneous (affine) deformations}
We first consider homogeneous deformations of the Rivlin square, for which we make the homogeneous ansatz
\begin{align}
\delta\vec{u}(\vec{X})=\tens{H}\cdot\vec{X}\qquad\Longrightarrow\qquad \delta\tens{F}=\tens{H},
\end{align}
where $\tens{H}$ is constant. Equation~\eqref{eq:deltaP_final} yields
\begin{align}
\delta\tens{P}=\mu\tens{S}-\frac{K}{2}\bigl(\tens{S}\teps-\teps\tens{S}\bigr)-\tau\tens{H}^\top-\delta p\,\tens{I},\label{eq:dP2}
\end{align}
where $\tens{S}=\tens{H}+\tens{H}^\top$. Since $\delta\tens{P}$ is thus constant, the incremental equilibrium is satisfied identically, and the boundary conditions yield
\begin{align}
\delta\tens{P}=\tzero.
\end{align}
The incremental incompressibility constraint \eqref{eq:incomp_u} gives $\tr{\tens{H}}=0$, so we write
\begin{align}
\tens{H}=\begin{pmatrix}
h & \gamma_1\\
\gamma_2 & -h
\end{pmatrix}.
\end{align}
Substituting into Eq.~\eqref{eq:dP2} and imposing $\delta\tens{P}=\tzero$ gives
\begin{subequations}
\begin{align}
(2\mu-\tau)h-K(\gamma_1+\gamma_2)-\delta p&=0,\\
-(2\mu-\tau)h+K(\gamma_1+\gamma_2)-\delta p&=0,\\
2Kh-\tau\gamma_2+\mu(\gamma_1+\gamma_2)&=0,\\
2Kh-\tau\gamma_1+\mu(\gamma_1+\gamma_2)&=0.
\end{align}
\end{subequations}
Adding the first two of these equations gives $\delta p=0$; taking the difference of the last two yields $\tau(\gamma_1-\gamma_2)=0$, so $\gamma_1=\gamma_2$ for $\tau\neq 0$. The remaining two independent equations then reduce to
\begin{align}
\begin{pmatrix}
2\mu-\tau & -2K\\
2K & 2\mu-\tau
\end{pmatrix}\begin{pmatrix}
h\\
\gamma_1
\end{pmatrix}=\begin{pmatrix}
0\\
0
\end{pmatrix}.
\end{align}
A non-trivial affine incremental solution therefore exists if and only if the determinant of the matrix on the left-hand side vanishes, which is if and only if
\begin{align}
(2\mu-\tau)^2+4K^2=0.
\label{eq:det_cond_incomp}
\end{align}
Since $\mu,\tau,K$ are real, this requires $\tau=2\mu$ and $K=0$. In particular, as noted in the main text, there is no affine solution for an odd square, $K\neq 0$, and so oddness breaks the pitchfork bifurcation of an incompressible even square~\cite{Overvelde2016} that occurs at $\tau=\tau_\ast=2\mu$.

\subsection{Inhomogeneous (plane-wave) deformations}
To analyse the possibility of inhomogeneous deformations, we introduce, again, a plane-wave ansatz
\begin{align}
\delta\vec{u}(\vec{X})=\vec{a}\,\e^{\i\vec{q}\cdot\vec{X}},\qquad \delta p(\vec{X})=\varpi\,\e^{\i\vec{q}\cdot\vec{X}},\label{eq:pwincomp}
\end{align}
with wave vector $\vec{q}$ and polarisation $\vec{a}$. Then $\delta\tens{F}=\i\vec{a}\otimes\vec{q}\e^{\i\vec{q}\cdot\vec{X}}$. The incompressibility constraint \eqref{eq:incomp_u} gives
\begin{align}
\vec{a}\cdot\vec{q}=0.\label{eq:incomppw}
\end{align}
In 2D, this means that there exists a scalar $\alpha$ such that
\begin{align}
\vec{a}=\alpha\,\teps\vec{q}.
\label{eq:a_alpha}
\end{align}
Because Eq.~\eqref{eq:deltaP_final} is homogeneous in $\delta\tens{F}$, we may drop the common factor $\i\e^{\i\vec{q}\cdot\vec{X}}$ and write
\begin{align}
\delta\tens{F}=\vec{a}\otimes\vec{q}.
\end{align}
As in the compressible case, the incremental equilibrium condition $\operatorname{Div}{\delta\tens{P}}=\vec{0}$ reduces to
\begin{align}
\delta\tens{P}\,\vec{q}=\vec{0}.\label{eq:DivPpw}
\end{align}
We now compute the various contributions in Eq.~\eqref{eq:deltaP_final}. First,\begin{subequations}
\begin{align}
\bigl(\delta\tens{F}+\delta\tens{F}^\top\bigr)\vec{q}&=(\vec{a}\otimes\vec{q}+\vec{q}\otimes\vec{a})\vec{q}=q^2\vec{a},
\end{align}
using Eq.~\eqref{eq:incomppw}, and where $q=\|\vec{q}\|$. Using Eq.~\eqref{eq:a_alpha}, we obtain, similarly,
\begin{align}
\bigl(\delta\tens{F}+\delta\tens{F}^\top\bigr)\teps\vec{q}=(\vec{a}\otimes\vec{q}+\vec{q}\otimes\vec{a})\teps\vec{q}=\alpha q^2\vec{q}.
\end{align}
Hence, noting that $\teps\vec{a}=-\alpha\vec{q}$ using Eq.~\eqref{eq:a_alpha},
\begin{align}
\bigl[\delta\tens{F}+\delta\tens{F}^\top,\teps\bigr]\,\vec{q}&=\bigl(\delta\tens{F}+\delta\tens{F}^\top\bigr)\teps\vec{q}-\teps\bigl(\delta\tens{F}+\delta\tens{F}^\top\bigr)\vec{q}\nonumber\\
&=\alpha q^2\vec{q}-\teps\bigl(q^2\vec{a}\bigr)=2\alpha q^2\vec{q},
\end{align}
Finally, using Eq.~\eqref{eq:incomppw} again,
\begin{align}
\delta\tens{F}^\top\vec{q}=(\vec{q}\otimes\vec{a})\vec{q}=\vec{0}.
\end{align}    
\end{subequations}
Substituting these expressions into Eq.~\eqref{eq:deltaP_final} gives
\begin{align}
\delta\tens{P}\,\vec{q}&=\mu q^2\vec{a}-\frac{K}{2}\bigl(2\alpha q^2\vec{q}\bigr)-\varpi\vec{q}=
\alpha q^2\mu\,\teps\vec{q}-\bigl(\alpha q^2K\!+\!\varpi\bigr)\vec{q}.
\end{align}
Now assume that $\vec{q}\neq\vec{0}$, so that the ansatz is inhomogeneous. Then $\vec{q}$ and $\teps\vec{q}$ are linearly independent so Eq.~\eqref{eq:DivPpw} implies
\begin{align}
\alpha q^2\mu=\alpha q^2K+\varpi=0.
\end{align}
Again, we expect $\mu\neq 0$, as discussed in the main text. Thus
\begin{align}
\alpha=0,\qquad \varpi=0.
\end{align}
Hence $\vec{a}=\vec{0}$ from Eq.~\eqref{eq:a_alpha}, and so $\delta\vec{u}=\vec{0}$ and $\delta p=0$ from Eqs.~\eqref{eq:pwincomp}, i.e., the plane-wave increment is necessarily trivial. In other words, the incompressible Rivlin square does not admit any nontrivial inhomogeneous incremental deformations.

\section{\uppercase{The Rivlin square for general constitutive relations}}
In proving, above, that the homogeneous bifurcation of the trivial square branch requires the parameter matching condition~\eqref{eq:para_match} for the compressible Rivlin square and is suppressed for the odd, incompressible Rivlin square, we assumed the minimal constitutive relations expressed by Eqs.~\eqref{eq:comp_T} and~\eqref{eq:incomp_T}, respectively. We now show that this conclusion does not depend on this assumption.
\subsection{The compressible case}
In the compressible case, it is easy to see that Eqs.~\eqref{eq:factorized_constraints_square} are replaced with
\begin{subequations}
\begin{align}
(\lambda_1-\lambda_2)\left[-t_1(\mathcal{I},\mathcal{J})+\frac{t_2(\mathcal{I},\mathcal{J})}{4}(\lambda_1+\lambda_2)^2\right]&=0, \\
(\lambda_1-\lambda_2)\left[-t_4(\mathcal{I},\mathcal{J})-\frac{t_3(\mathcal{I},\mathcal{J})}{2}(\lambda_1+\lambda_2)^2\right]&=0.
\end{align}
\end{subequations}
Away from the square branch, both factors in square brackets vanish. The bifurcation of the square branch occurs at ${\lambda_1=\lambda_2=\lambda_\ast}$, and hence $\mathcal{I}=2\lambda_\ast$, $\mathcal{J}=\lambda_\ast^2$. We thus obtain the two simultaneous equations 
\begin{subequations}
\begin{align}
t_1\bigl(2\lambda_\ast,\lambda_\ast^2\bigr)-t_2\bigl(2\lambda_\ast,\lambda_\ast^2\bigr)\lambda_\ast^2&=0,\\
t_4\bigl(2\lambda_\ast,\lambda_\ast^2\bigr)+2t_3\bigl(2\lambda_\ast,\lambda_\ast^2\bigr)\lambda_\ast^2&=0
\end{align}
\end{subequations}
for the critical stretch $\lambda_\ast$. Solving one of these determines (one of possibly many values of) $\lambda_\ast$. The other equation is then not satisfied in general, and therefore expresses a parameter matching condition for the bifurcation.

\subsection{The incompressible case}
In the incompressible case, a similar argument leads to the conditions
\begin{subequations}
\begin{align}
    (\lambda_1-\lambda_2)\left[p + t_2(\mathcal{I})\right] = 0, \\
    (\lambda_1-\lambda_2)(\lambda_1 + \lambda_2)^2t_3(\mathcal{I}) = 0.
\end{align}
\end{subequations}
Because the square does not deform on the trivial branch by incompressibility, $\lambda_1=\lambda_2=1$ and hence $\mathcal{I}=2$ at the bifurcation point. Hence
\begin{align}
p + t_2(2)=t_3(2) = 0.
\end{align}
However, $t_3(2)=-K$ from Eq.~\eqref{eq:match_shear_main}, so the square bifurcates if and only if the material is even.

\section{\uppercase{3D isotropic Cauchy elasticity}}
As discussed in the main text, the Cauchy stress of an incompressible isotropic Cauchy elastic solid admits the 3D representation~\cite{yavari2025}
\begin{align}
\tens T = -p\tens{I} + \alpha(\mathcal{I}_1, \mathcal{I}_2)\tens{B} + \beta(\mathcal{I}_1, \mathcal{I}_2)\tens{B}^2,
\label{eq:T_3D}
\end{align}
where $\mathcal{I}_1 = \tr \tens{B}$, $\mathcal{I}_2 = \left(\mathcal{I}_1^2-\tr{\tens{B}^2}\right)/2$. As announced in the main text, we truncate expansions of $\alpha,\beta$ around the undeformed state $\mathcal{I}_1=3$, $\mathcal{I}_2=3$ to linear order, so that
\begin{subequations}\label{eq:ab}
\begin{align}
\alpha(\mathcal{I}_1,\mathcal{I}_2)&=\alpha_{00}+\alpha_{10}(\mathcal{I}_1-3)+\alpha_{01}(\mathcal{I}_2-3),\\
\beta(\mathcal{I}_1,\mathcal{I}_2)&=\beta_{00}+\beta_{10}(\mathcal{I}_1-3)+\beta_{01}(\mathcal{I}_2-3).
\end{align}
\end{subequations}
\subsection*{Triaxial bifurcations of the Rivlin cube}
We now analyse the possibility of a triaxial bifurcation from the trivial branch of the Rivlin cube. As expressed by Eq.~(15) of the main text, on such a triaxial branch, 
\begin{align}
\mathcal{C}=\alpha+\beta r=0,
\end{align}
where $r =\lambda_1^2+\lambda_2^2+\lambda_3^2+\lambda_1\lambda_2+\lambda_2\lambda_3+\lambda_3\lambda_1$. We determine this triaxial branch close to $\lambda_1=\lambda_2=\lambda_3=1$ by expanding, without loss of generality,
\begin{subequations}
\begin{align}
\lambda_1=1+\epsilon\Lambda,
\end{align}
where $\epsilon\ll 1$ and $\Lambda$ is a control parameter along the branch, which is not zero away from the bifurcation point. We further expand
\begin{align}
\lambda_2=1+\epsilon\Lambda_2^{(1)}+\epsilon^2\Lambda_2^{(2)}+\epsilon^3\Lambda_2^{(3)}+\epsilon^4\Lambda_2^{(4)}+\mathcal O\bigl(\epsilon^5\bigr),\\
\lambda_3=1+\epsilon\Lambda_3^{(1)}+\epsilon^2\Lambda_3^{(2)}+\epsilon^3\Lambda_3^{(3)}+\epsilon^4\Lambda_3^{(4)}+\mathcal O\bigl(\epsilon^5\bigr),
\label{eq:lx}
\end{align}
\end{subequations}
Using \textsc{Mathematica}, the incompressibility constraint $\lambda_1\lambda_2\lambda_3=1$ gives, at leading order,
\begin{subequations}
\begin{align}
\Lambda+\Lambda_2^{(1)}+\Lambda_3^{(1)}=0,\label{eq:ic1}
\end{align}
which we use to eliminate $\smash{\Lambda_3^{(1)}}$. Continuing this expansion, we find
\begin{align}
-\Lambda^2-\Lambda\Lambda_2^{(1)}-\bigl(\Lambda_2^{\smash{(1)}}\bigr)^2+\Lambda_2^{(2)}+\Lambda_3^{(2)}=0,\label{eq:ic2}
\end{align}
which allows us to eliminate $\smash{\Lambda_3^{(2)}}$, too. We then obtain
\begin{align}
&\Lambda^3+\Lambda^2\Lambda_2^{(1)}+\Lambda\bigl(\Lambda_2^{\smash{(1)}}\bigr)^2+\bigl(\Lambda_2^{\smash{(1)}}\bigr)^3-\Lambda\Lambda_2^{(2)}-2\Lambda_2^{(1)}\Lambda_2^{(2)}\nonumber\\
&\qquad+\Lambda_2^{(3)}+\Lambda_3^{(3)}=0,
\end{align}
\end{subequations}
which we can use to eliminate $\smash{\Lambda_3^{(3)}}$. With this, we expand, first
\begin{align}
\mathcal{C}&=\mathcal{C}_0+\epsilon^2\mathcal{C}_2\left[\Lambda^2+\Lambda\Lambda_2^{\smash{(1)}}+\bigl(\Lambda_2^{\smash{(1)}}\bigr)^2\right]+\mathcal O\bigl(\epsilon^3\bigr),
\end{align}
where
\begin{subequations}
\begin{align}
\mathcal{C}_0&=\alpha_{00}+6\beta_{00},\\
\mathcal{C}_2&=4(\alpha_{01}+\alpha_{10})+5\beta_{00}+24(\beta_{01}+\beta_{10}).
\end{align}
\end{subequations}
Since $\smash{\Lambda^2+\Lambda\Lambda_2^{\smash{(1)}}+\bigl(\Lambda_2^{\smash{(1)}}\bigr)^2}>0$ for all real $\Lambda,\smash{\Lambda_2^{\smash{(1)}}}$, $\mathcal{C}=0$ requires
\begin{align}
\mathcal{C}_0=\mathcal{C}_2=0.\label{eq:c0c2}
\end{align}
Imposing these conditions, the expansion of $\mathcal{C}$ becomes
\begin{align}
\mathcal{C}&=\epsilon^3\mathcal{C}_3\Lambda\Lambda_2^{\smash{(1)}}\bigl(\Lambda+\Lambda_2^{\smash{(1)}}\bigr)+\mathcal O\bigl(\epsilon^4\bigr),\label{eq:C3exp}
\end{align}
where
\begin{align}
\mathcal{C}_3=\dfrac{2}{5}\bigl[17\alpha_{01}-3(\alpha_{10}-34\beta_{01}+6\beta_{10})\bigr].
\end{align}
We claim that Eqs.~\eqref{eq:c0c2} provide necessary and sufficient conditions for a triaxial bifurcation off the trivial branch. We will not discuss the pathological case $\mathcal{C}_3=0$. Now $\Lambda\neq0$ away from the trivial branch by definition, so Eq.~\eqref{eq:C3exp} implies that $\smash{\Lambda_2^{(1)}}=0$ or $\smash{\Lambda+\Lambda_2^{(1)}=-\Lambda_3^{(1)}}=0$, using Eq.~\eqref{eq:ic1}. Without loss of generality, we choose $\smash{\Lambda_3^{(1)}}=0$, whence ${\smash{\Lambda_2^{(1)}}=-\Lambda}$. 

At this stage, we continue the expansion of the incompressibility condition to order $\mathcal{O}\bigl(\epsilon^4\bigr)$ to obtain
\begin{align}
\Lambda_3^{(4)}=\Lambda^4-2\Lambda^2\Lambda_2^{(2)}+\bigl(\Lambda_2^{\smash{(2)}}\bigr)^2-\Lambda\Lambda_2^{(3)}-\Lambda_2^{(4)}.
\end{align}
Imposing $\mathcal{C}=0$ to order $\mathcal{O}\bigl(\epsilon^4\bigr)$ then yields a linear equation for $\smash{\Lambda_2^{(2)}}$. With this, Eq.~\eqref{eq:ic2} determines $\smash{\Lambda_3^{(2)}}$, too, giving a parametric expression of the triaxial branch to order $\mathcal O\bigl(\epsilon^2\bigr)$,
\begin{subequations}
\begin{align}
\lambda_1&=1+\epsilon\Lambda,\\
\lambda_2&=1-\epsilon\Lambda+\epsilon^2\Lambda^2\dfrac{5(7\alpha_{01}-\alpha_{10}+62\beta_{01}+14\beta_{10})}{2(17\alpha_{01}-3\alpha_{10}+102\beta_{01}-18\beta_{10})}\nonumber\\
&\qquad+\mathcal O\bigl(\epsilon^3\bigr),\\
\lambda_3&=1-\epsilon^2\Lambda^2\dfrac{\alpha_{01}+\alpha_{10}+106(\beta_{01}+\beta_{10})}{34\alpha_{01}-6(\alpha_{10}-34\beta_{01}+6\beta_{10})}+\mathcal O\bigl(\epsilon^3\bigr).
\end{align}
This assumes that the necessary and sufficient conditions~\eqref{eq:c0c2} hold. From this, we can also obtain an expansion of $\tau$ to order $\mathcal O\bigl(\epsilon^2\bigr)$, but this is of no further interest.
\end{subequations}
\bibliography{main.bib}